\documentstyle[preprint,eqsecnum,aps]{revtex}
\begin{document}
\draft
\preprint{P-94-05-033}

\title{
\parbox{6in}{\normalsize
\vspace{-.5in}ICHEP94~Ref.~gls0167 \\
Submitted~to~Pa-09, Pl-10 \\
}  \\
\large {\bf New Measurements of CP Violation Parameters \\
as Tests of CPT in $K$ Meson Decay\cite{thestudents} }
\footnote {presented at the {\rm $27^{\rm th}$}
International Conference on High Energy Physics, Glasgow, July, 1994}}

\author{G.D. Gollin}
\address{Dept. of Physics, University of Illinois at Urbana-Champaign,
Urbana, Illinois 61801-3080}
\author{W.P. Hogan}
\address{Department of Physics,
Rutgers University, Piscataway, New Jersey 08855 }
%
%
\author{{\bf {(FNAL E773 collaboration)}}}
\address{\phantom {leave a blank space}}

\date{June 29, 1994}

\maketitle
%
%
%
%
\begin{abstract}
Using a technique which employs a pair of solid scintillator
regenerators, the E773 collaboration has measured several CP violation
parameters in $K$
meson decay at Fermilab.
We report new results for the phase of
$\eta_{+-}$, the $K_L-K_S$ mass difference,
the $K_S$ lifetime,
and the phase difference
$Arg(\eta_{00})-Arg(\eta_{+-})$
in $K \rightarrow \pi \pi$ decay.
In addition, we report a measurement of the magnitude and phase of
$\eta_{+-\gamma}$ in
$K \rightarrow \pi^+ \pi^- \gamma$ decay.
Our preliminary results are compared with theoretical expectations based on
CPT symmetry.
\\
\\
\\
\end{abstract}

%
%
%
\pacs{PACS numbers: 10.13.20.Eb, 10.14.40.Aq}
\tableofcontents
%
\narrowtext
\section{Introduction}
\label{sec:introduction}
%
%
Local quantum field theories are automatically invariant
under the combined operations of
charge conjugation, parity, and time reversal: CPT \cite{streater}.
However, generalizations of quantum mechanics which include gravity
might allow pure states to evolve into mixed states \cite{hawking75}.
This sort of unusual (Planck-scale) dynamics could lead to
CPT-violating effects \cite{ellis84,peskin94,ellis92a,ellis92b}
such as the existence of
small differences in
particle-antiparticle masses and lifetimes.  The magnitude of this
CPT-noninvariance might be proportional to the ratio of some low energy scale
to the Planck mass (Ref.~\cite{ellis92b}).
For example, one might have
\begin{eqnarray}
\frac{m_p - m_{\overline p}}{m_p}\ \ \   \sim \ \ \frac{m_p}{m_{Planck}}
\ \ \ = \ \ 7.7 \times 10^{-20} \nonumber
\end{eqnarray}
for the proton-antiproton mass difference, and
\begin{eqnarray}
{m_{K^0} - m_{\overline {K^0}} \over m_{K}}\ \ \sim\ \ {m_{K} \over
m_{Planck}} \ \ = \ \ 4.1 \times 10^{-20} \nonumber
\end{eqnarray}
for the neutral $K$ system.

%
The 90\% confidence level limits on the proton-antiproton and
electron-positron
fractional mass differences are approximately $4 \times 10^{-8}$,
twelve orders of magnitude larger than the ``interesting" region \cite{PDG}.
The experimental situation is considerably brighter, however, in the $K^0$
sector.
Let us define
\begin{eqnarray}
&&\eta _{+-}\ \equiv\ {Amp (K_L \rightarrow \pi^+\pi^-)\over
Amp(K_S \rightarrow \pi^+\pi^-)}\ , \ \ \ \eta_{00}\ \equiv \
{Amp(K_L\rightarrow\pi^0\pi^0) \over Amp(K_S\rightarrow
\pi^0\pi^0)}\ , \nonumber \\
&&\phi_{+-}\ \equiv \ Arg(\eta_{+-})\ ,\ \ \ \ \phi_{00}\ \equiv \
Arg(\eta_{00})\ , \ \ \ \Delta\phi\ \equiv\ \phi_{00}\ -\ \phi_{+-}\ .
\nonumber
\end{eqnarray}
It can be shown\cite{carosi90} that
\[
{m_{K^0} - m_{\overline {K^0}} \over m_{K}} \ \ \approx\ \
\left( {{\Delta m} \over {m_K }} \right)
\sqrt{2} \, \vert\eta_{+-} \vert \,\,
tan(\phi_{+-} - \phi_\epsilon + {\Delta\phi\over 3})\ .
\]
Here,
$\Delta m = 0.5286 \times 10^{10} \hbar s^{-1}c^{-2}
= 3.4793 \times 10^{-6}eV/c^2$
is the $K_L - K_S$ mass difference.
The small value for $\Delta m$ provides
considerable ``leverage"
in testing CPT since $\Delta m / m_K = 6.99 \times 10^{-15}$. 
Recently published values\cite{gibbons1} yield
$\vert m_{K^0} - m_{\overline {K^0}} \vert / m_{K} \alt 2.5 \times 10^{-18}$,
less than two orders of magnitude away from
the domain in which Planck-scale physics might play a role.

%
In Fermilab
Experiment 773 we measured $\phi_{+-}$, $\phi_{\epsilon}$, and
$\Delta \phi$ by studying
decays of neutral $K$ mesons into $\pi^0\pi^0$ and $\pi^+\pi^-$ final states.
In addition to our CPT-related $\pi\pi$
results, we also report an improved measurement
of
$\eta_{+-\gamma}$, a CP-violation parameter
in $K_L \rightarrow \pi^+\pi^-\gamma$ decays.
We determined the phases of decay amplitudes by observing
the time-dependent
interference between $K_S$ and $K_L$ decays
after a pair of $K_L$ beams passed through
regenerators.
The detector was a reconfigured version of the E731 spectrometer
(Ref.~\cite{gibbons1})
which had been
used to measure $Re(\epsilon'/\epsilon)$ and other parameters of the neutral
$K$
system.

E773 wrote physics-quality data for about two months,
beginning in late July, 1991.  During this time the experiment recorded
$\sim$400 million triggers on nine hundred 8~mm cassette data tapes.
We present preliminary results from analysis of these
data.
In
Sec.~\ref{sec:phenomenology} we
review relevant neutral $K$ phenomenology.
In Secs.\  \ref{sec:technique} and \ref{sec:detector_calibration}
we describe the experimental technique and instrumentation used in E773.
In Secs.~\ref{sec:chgreconstruction} and \ref{sec:neureconstruction}
we discuss reconstruction of
candidate $K \rightarrow \pi\pi$ and $K \rightarrow \pi\pi\gamma$
events. Simulation of the detector is discussed
in Sec.~\ref{sec:montecarlo}.
We describe analysis and fits to our data in Sec.~\ref{sec:analysis}.
Our conclusions are presented in Sec.~\ref{sec:conclusions}.
These results comprise the thesis
work of R.A. Briere and B. Schwingenheuer from the University of
Chicago, and J.N. Matthews from Rutgers University.

\section{$K$ Meson Phenomenology}
\label{sec:phenomenology}
%
%
\subsection{CP and $\pi\pi$ decays}
\label{sec:CP_pipi_phenomenology}
The major source of CP violation in the neutral $K$ system\cite{gibbons2,NA31}
is
the time-reversal asymmetry in the transition rate between $K^0$ and
$\overline{K^0}$:
$\Gamma (\overline{K^0} \rightarrow K^0) >
\Gamma (K^0 \rightarrow \overline{K^0})$.
Consequently, the
eigenstates of the neutral $K$ system which
do not mix in vacuum
will both
contain an excess of $K^0$ relative to $\overline{K^0}$.  We define
\[
\vert K_S \rangle \equiv {{(1+\epsilon)\vert K^0 \rangle +
(1-\epsilon)\vert \overline{K^0} \rangle} \over
{[2(1+\vert \epsilon \vert^2)]^{1/2}}} \ ,
\ \ \vert K_L \rangle \equiv {{(1+\epsilon)\vert K^0 \rangle -
(1-\epsilon)\vert \overline{K^0} \rangle} \over
{[2(1+\vert \epsilon \vert^2)]^{1/2}}}\ .
\]
The phase of $\epsilon$ (Ref.~\cite{PDG}) is
$$\phi_\epsilon\  \approx \  tan^{-1}\left({2\Delta mc^2 \tau_s
\over \hbar}\right) = (43.33\pm 0.14)^\circ \ ,$$
so that $Re(\epsilon)$ is positive.
The $K_S$ lifetime is $\tau_S = 0.8922 \times 10^{-10}s$\cite{gibbons1}.
The CP eigenstates are
\[
\vert K_1 \rangle \equiv {{\vert K^0 \rangle +
\vert \overline{K^0} \rangle} \over {\sqrt{2}}} \ ,
\ \ \vert K_2 \rangle \equiv {{\vert K^0 \rangle -
\vert \overline{K^0} \rangle} \over {\sqrt{2}}}\ .
\]
Neglecting normalizations, we have
$\vert K_1 \rangle \sim \vert K_S \rangle - \epsilon \vert K_L \rangle$
and $\vert K_2 \rangle \sim \vert K_L \rangle - \epsilon \vert K_S \rangle$.
A beam which is initially pure $K_2$ will evolve into a mixture of $K_2$ and
$K_1$ as the $K_S$ component of the original $K_2$ state decays.
This CP-violating mixing $K_2 \leftrightarrow K_1$, combined with the
(CP-allowed) decay $K_1 \rightarrow \pi \pi$, provides the dominant
contribution to the $K_L \rightarrow \pi \pi$ decay amplitude.

An additional source of CP violation in $K_L \rightarrow \pi \pi$ decay is
possible if $Amp(K_2 \rightarrow \pi \pi)$ is nonzero.  Since
$\vert K_2 \rangle \sim \vert K^0 \rangle - \vert {\overline {K^0}} \rangle$,
this {\it direct} CP violation is only possible if
$Amp(K^0 \rightarrow \pi \pi) \neq Amp({\overline {K^0}} \rightarrow \pi \pi)$.
More specifically, we define the isospin 0 and 2 decay amplitudes $A_0$,
${\overline {A_0}}$,
$A_2$,
${\overline {A_2}}$
so that
\begin{eqnarray}
&&\exp^{i\delta_0} A_0 = \langle (\pi\pi)_{I=0}\vert T \vert K^0 \rangle
\ ,\ \ \exp^{i\delta_0} {\overline {A_0}} =
\langle (\pi\pi)_{I=0}\vert T \vert {\overline{K^0}} \rangle
\nonumber \\
&&\exp^{i\delta_2} A_2 = \langle (\pi\pi)_{I=2}\vert T \vert K^0 \rangle
\ ,\ \ \exp^{i\delta_2} {\overline {A_2}} =
\langle (\pi\pi)_{I=2}\vert T \vert {\overline{K^0}} \rangle \ .
\nonumber
\end{eqnarray}
We find that
\[
\epsilon' \equiv {1 \over \sqrt{2}} \exp^{i(\delta_2 - \delta_0)}
{ {(A_2 - {\overline {A_2}})} \over {(A_0 + {\overline {A_0}})} }
\]
must be different from zero
for there to be direct CP violation in $K$ decay.
Here, $\delta_0$ and $\delta_2$ are the
(measured)
isospin 0 and 2 $\pi\pi$ scattering phase shifts\cite{devlin} at the
$K^0$ mass. CPT invariance requires ${\overline {A_I}}={A_I}^*$, so the
phase of $\epsilon'$ is
$\phi_{\epsilon'} = \delta_2 - \delta_0 + 90^\circ \approx (47 \pm 5)^\circ$
(Ref.~\cite{PDG}).
It may seem surprising
that the phase of $\epsilon'$ is well known, even though its
magnitude remains an open question (Ref.~\cite{gibbons1,NA31}).
By coincidence, the phases of $\epsilon$ and $\epsilon'$
are nearly equal.
If we assume CPT invariance, we can write $\eta_{+-}$ and
$\eta_{00}$ in terms
of $\epsilon$ and $\epsilon'$ as
\begin{equation}
\eta_{+-} \approx \epsilon + \epsilon' \ ,\ \ \
\eta_{00} \approx \epsilon - 2\epsilon'.      \label{etasdef}
\end{equation}

\subsection{CP and $\pi\pi\gamma$ decays}
\label{sec:CP_pipigamma_phenomenology}
CP considerations play a similar role in $K \rightarrow \pi^+\pi^-\gamma$
decays.
The $K_S \rightarrow \pi\pi\gamma$ rate is dominated by an inner bremsstrahlung
amplitude\cite{taureg}
in which one of the final-state pions radiates a photon.
In $K_L$ decays, the inner bremsstrahlung contribution
is CP-suppressed, permitting observation of the CP-allowed direct emission
process\cite{ramberg1}.  As is the case for $\pi\pi$ decays, the CP-violating
mixing
$K_2 \leftrightarrow K_1$, combined with the
(CP-allowed) decay $K_1 \rightarrow \pi \pi \gamma$, contributes to the
$\pi \pi \gamma$ decay rate in a $K_L$ beam\cite{ramberg2}.
We define the ratio of the inner bremsstrahlung decay amplitudes:
\[
\eta_{+-\gamma}\ \equiv
\ {Amp(K_L \rightarrow \pi^+\pi^-\gamma)_{\rm IB}\over
Amp(K_S \rightarrow \pi^+\pi^-\gamma)_{\rm IB}}\ .
\]
Unless there is an unusually large amount of direct CP violation in
$K_L \rightarrow \pi^+\pi^-\gamma$ decays, one expects
$\eta_{+-\gamma}\approx\eta_{+-}\approx\epsilon$. We
only observe interference between the inner bremsstrahlung
amplitudes;
the $K_S$ direct emission amplitude is too
small to be detected in our data.

\subsection{CPT}
\label{sec:CPT_pipi_phenomenology}
Several different avenues for CPT violation suggest
themselves.  One is a CP-violating, T-conserving mixing of
$K^0$ and ${\overline{K^0}}$
to produce the mass eigenstates $K_L,\ K_S$.
This corresponds to a situation where the $K_S$
contains, for example, an excess of $K^0$
while the $K_L$ contains an excess of ${\overline{K^0}}$.
We can describe this by including
the CPT violation parameter $\Delta$
in our description of $K_L,\ K_S$ as follows:
\begin{eqnarray}
\vert K_S \rangle\ \sim\ {{(1+\epsilon+\Delta)\vert K^0 \rangle\ \ +
\ \ (1-\epsilon-\Delta)\vert \overline{K^0} \rangle}} \phantom{\ .}\nonumber\\
\vert K_L \rangle\ \sim\ {{(1+\epsilon-\Delta)\vert K^0 \rangle\ \ -
\ \ (1-\epsilon+\Delta)\vert \overline{K^0} \rangle}}\ . \nonumber
\end{eqnarray}
Note the sign switch between $\epsilon$ and $\Delta$ in the $K_L$ expression.
The component of $\Delta$
which is perpendicular to $\epsilon$ in the complex plane
corresponds to a  $K^0 -  \overline{K^0}$
mass difference, while the component parallel to $\epsilon$
corresponds to a
lifetime difference\cite{cronin}.
In terms of $\epsilon$, $\epsilon'$, and $\Delta$,
\begin{equation}
\eta _{+-}\ =\ \epsilon + \epsilon' - \Delta\ ,\ \ \ \
\eta _{00}\ =\ \epsilon - 2\epsilon' - \Delta\ .  \label{etadelta}
\end{equation}
A nonzero value of $\Delta$ will shift $\eta_{00}$ and  $\eta_{+-}$
in the same direction in
the complex plane.
{}From Eq.~\ref{etadelta},
we find $\Delta = \epsilon -(2 \eta_{+-} +\eta_{00})/3$.

Another possible route to CPT violation is through an unusual
relationship among the various
$K \rightarrow \pi\pi$
decay amplitudes.
Because
$A_I - {A_I}^* = 2iIm(A_I)$ and $A_I + {A_I}^* = 2Re(A_I)$,
the phase of $\epsilon'$ would be shifted
from its value of $(47 \pm 5)^\circ$
by a CPT-disallowed relationship such as
${\overline {A_I}} \neq {A_I}^*$.
This would split $\eta_{00}$ and  $\eta_{+-}$
apart in
the complex plane.
Other terms associated with CPT violation
in the decay amplitudes would shift
$\eta_{00}$ and  $\eta_{+-}$ in the same direction, much as was the
case with $\Delta$.
{}From Eq.~\ref{etadelta}, we see that
$\epsilon' \approx (\eta_{+-} -  \eta_{00})/3$,
even in the presence of additional (CPT violating)
terms which are common to both
$\eta_{00}$ and  $\eta_{+-}$.

\subsection{$K_S$ regeneration}
\label{sec:ksregeneration}
We extracted phase information about the $\eta's$ by measuring
the interference between the
$K_L$ and $K_S$ decay amplitudes in our neutral beam.
Since the E773 detector was many $K_S$ lifetimes downstream of the
production
target, we reintroduced a small $K_S$ component into our
twin $K_L$ beams by passing both beams through regenerators made of
plastic scintillator.

Because the forward
elastic scattering amplitudes $f(k)$ and ${\overline {f}}(k)$
for $K^0$ and ${\overline {K^0}}$ with wavenumber $k$ are unequal,
a $K_L$ entering a block of material will evolve into a mixture
of $K_L$ and $K_S$.
The ratio of the $K_S$ and $K_L$ amplitudes a small distance
$\Delta z$ inside a regenerator will be
\[
\rho \equiv {{A_S(\Delta z)} \over {A_L(\Delta z)}} \approx iN\pi
\left({{f - {\overline f}}\over {k}}\right)\Delta z \ .
\]
Here, $N$ is the number of scatterers per unit volume.
This {\it coherent regeneration amplitude} $\rho$
results from constructive interference among all the outgoing spherical waves
produced at each of the scattering sites in the regenerator.
For a finite length regenerator, $\rho$ includes an additional
geometric factor associated with the $K_S$ lifetime and the $K_L - K_S$
mass difference.\cite{kleinknecht}
The momentum-dependence of $\vert (f - {\overline f})/k \vert$
has been measured
experimentally\cite{gibbonsthesis} and is in
agreement with the
Regge theory prediction\cite{gilman} that
\begin{equation}
\left({{f - {\overline f}} \over k}\right) \ \propto
\ p^{\alpha}e^{-i \pi (2+\alpha)/2}\ .
\label{regge}
\end{equation}
The connection between $\alpha$ and the phase of $f-{\overline f}$ comes
from analyticity, and is independent of Regge theory.

A kaon which collides with a single nucleus
will
scatter into a state containing a different combination
of $K^0$ and ${\overline {K^0}}$.
Consequently,
elastic scattering is regenerative:
a pure $K_L$
will scatter into a mixed
$K_L$, $K_S$ state.
Using the optical theorem, one can show that
the contribution to the $K_S$ amplitude from
elastic scattering
interferes
destructively
with the scattered kaon's $K_S$ amplitude which arises from
coherent regeneration.  This cancellation is nearly perfect in a
regenerator of two interaction lengths.

We were interested in studying the interference between the
(coherently regenerated)
$K_S$ and $K_L$ decay amplitudes.
Since
E773's regenerators were shorter than two interaction lengths,
we made small
corrections for contamination
in our coherently regenerated $K$ sample from
``diffraction regeneration" associated with single (and multiple)
elastic scattering.
Regeneration is discussed more fully in the Appendix.

\subsection{Decay rate downstream of a regenerator}
\label{sec:decayrate}
The $K_S$ and $K_L$ $\pi\pi$ decay amplitudes will interfere downstream of
a regenerator.
Because of the differences in the
$K_L$, $K_S$ masses and decay rates,
this interference evolves with proper time $\tau$.
For a state
$\vert K \rangle \sim
\vert K_L \rangle +\rho \vert K_S \rangle$,
$$Amp(K \rightarrow \pi \pi) =
Amp(K_L \rightarrow \pi \pi) e^{-im_Lc^2\tau/\hbar - \Gamma_L \tau/2}
+
\rho Amp(K_S \rightarrow \pi \pi) e^{-im_Sc^2\tau/\hbar - \Gamma_S \tau/2}\ ,
$$
where $\Gamma_L$, $\Gamma_S$ are the $K_L$, $K_S$ decay rates.
The $\pi\pi$ decay rate
downstream of the regenerator
is proportional to
$$|\rho|^{2}e^{-\Gamma_S\tau} + |\eta|^{2}e^{-\Gamma_L\tau} +
2|\rho\eta|e^{-(\Gamma_L+\Gamma_S)\tau/2} \cos (\Delta mc^2\tau/\hbar
 + \phi_\rho-
\phi_\eta)\ .$$
The phase
of $\rho$
is approximately -32$^\circ$, though it
varies with energy by several degrees
since the $K_L$ phase advances with respect to the
$K_S$ phase as a kaon travels through a regenerator.
$\Gamma_S$ is $1.12 \times 10^{10}$s$^{-1}$, $\Gamma_L$ is
$1.93 \times 10^7$s$^{-1}$,  and $\Delta mc^2/\hbar$ is
$30.6^\circ/10^{-10}$s, about
$5^\circ$ per meter for 100~GeV kaons.
The last term in the decay rate,
the interference
between $K_L$ and $K_S$ decay amplitudes, contains information
about $\phi_{00}$ or $\phi_{+-}$.  Defining
$\Gamma_{+-}$ as the $K \rightarrow \pi^+\pi^-$
decay rate, $\Gamma_{00}$ as the $K \rightarrow \pi^0\pi^0$
decay rate, and $\Delta\phi$ as $\phi_{00}-
\phi_{+-}$, one may write
\begin{eqnarray}
\Gamma_{{\scriptscriptstyle +-}} &\ \sim\ &
|\rho|^{2} e^{-1.12\tau} +\ |\eta_{+-}|^2  e^{-.00193\tau} +
\ 2|\rho\eta_{+-}| e^{-.56\tau} \cos (30.6\tau-77)^\circ \nonumber \\
\Gamma_{00}\, &\ \sim\ & |\rho|^{2} e^{-1.12\tau}
+\ |\eta_{00}|^2  e^{-.00193\tau} +
\ 2|\rho\eta_{00}| e^{-.56\tau} \cos(30.6\tau - 77 - \Delta\phi)^\circ
\nonumber
\end{eqnarray}
where $\tau$ is in units of $10^{-10}$ seconds
and the arguments of the cosines are in degrees. The interference term in
$\Gamma_{00}$ is most sensitive to changes in
$\Delta\phi$ when the cosine's
argument is near -90$^\circ$, corresponding to $\tau \sim 0$. For
$\tau \sim 2.5 \times 10^{-10}$s
the argument of the cosine is near zero
and the $\pi^0\pi^0$ decay rate is insensitive to small changes
in $\Delta\phi$.
%
%

\section{Experimental Technique}
\label{sec:technique}

If experimental acceptance and detection efficiency are known
with sufficient accuracy, $\Delta\phi$, $\phi_{+-}$, and $Arg(\epsilon)$
can be determined
in a single beam through a measurement of the proper time
evolution of  $\Gamma_{00}$ and $\Gamma_{+-}$.
However,
resolution smearing,
diffraction regeneration,
and experimental acceptance influence
the reconstruction of
charged and neutral final states in different ways.
As a result, the E773 apparatus was designed conservatively
to minimize possible systematic errors in the
measurement of $\Delta \phi$
associated with acceptance modeling, energy calibration,
and imperfect knowledge
of physics parameters such as $\epsilon'$.
(We subsequently found that our ability to model the
behavior of the detector was adequate to make single-beam
measurements.)

\subsection{Double beam technique}
\label{sec:double_beam}
We employed a double-beam technique
similar to that used by E731 in its measurement (Ref.~\cite{gibbons2})
of $\epsilon'$.
One $K_L$ beam passed through
a thin regenerator while
the other traversed a thick regenerator about eleven meters further
upstream.
This separation
was chosen to make the $\pi\pi$ decay rate
in the upstream regenerator's beam
insensitive to $\Delta\phi$ after the beam had traveled
about a dozen meters downstream of the
upstream regenerator, corresponding to
proper times $\sim 2.5 \times 10^{-10}s$.
The two beams were separated vertically; individual
kaons passed through
only one of the two regenerators.
Data were recorded simultaneously for $\pi^0\pi^0$ and
$\pi^+\pi^-$ decays
in both beams. The regenerators switched beams after each accelerator
spill, approximately once per minute.
Various systematic effects associated with detection efficiency,
kaon flux, and acceptance cancel when rates in the two beams are compared.

\subsection{Choice of regenerator configuration}
\label{sec:reglen}
A conservative analysis of the data could be performed by forming a double
ratio
of the decay rates in the two beams for the neutral and charged decay modes.
For a given value of kaon energy
$E$ and decay position $z$, we define
$$R(z,E) \equiv {\Gamma_{00}^{up}(z,E) \over \Gamma_{00}^{down}(z,E)} \bigg/
{\Gamma_{+-}^{up}(z,E) \over \Gamma_{+-}^{down}(z,E)}\ .$$
Here, ``up" and ``down" refer to the beams
containing regenerators in the upstream and downstream
positions.  Binned in $E$ and $z$, $R$ is insensitive to
differences in $\pi^0\pi^0$  and $\pi^+\pi^-$ acceptance and
reconstruction efficiency.
For
$z$ values downstream of the thin regenerator,
$\Gamma_{00}^{up}(z,E)$ is nearly independent
of $\Delta \phi$ for typical kaon energies  while
$\Gamma_{00}^{down}(z,E)$ changes by about one percent per
one degree change in $\Delta \phi$.

A regenerator which
is approximately one interaction length in thickness
gives the largest amount of
$K_L/K_S$ interference, since $\rho$ is proportional to regenerator
length, while the fraction of beam surviving passage through the
regenerator falls exponentially with length.
Most of the coherent $K_S$ flux in the ``upstream" beam,
proportional to $\rho^2$, will have decayed before traveling as far
as the downstream regenerator.
A judicious choice for the thickness of the
downstream regenerator will produce a decay distribution in
this beam with similar time dependence,
except for the proper-time offset in the
interference cosine's argument.
By choosing regenerator lengths which give similar proper time
spectra, the importance of systematic errors caused by
resolution
effects is reduced.
In particular, the double ratio $R(z,E)$ will be close to unity for
a wide range of decay positions and energies,
and will vary from unity by typically 1\% per degree of phase
difference between $\eta_{00}$ and $\eta_{+-}$.
The optimum regenerator thicknesses proved to be 0.4 interaction
lengths for the downstream and 1.2 interaction lengths for the
upstream regenerators.

A drawback of a double ratio analysis is that it
discards events with decay positions
upstream of the downstream regenerator, sacrificing statistical
power in order to reduce systematic uncertainties.
With adequate modeling of the signal and backgrounds, it is possible to
extract additional information about the $\eta$ phases from events ignored
by the double ratio technique.
We found that we were able to understand (and simulate) the behavior of
the E773 detector with sufficient accuracy
to overcome the systematic effects that were of concern at the
time the detector was configured.
Consequently, we report results from
the higher precision analysis which includes data from the region
between the regenerators.

\section{E773 Detector and Calibration}
\label{sec:detector_calibration}
The E773 detector\cite{p773} was designed
to determine decay positions and momenta of
$K$ mesons decaying into $2\pi$, $\pi \pi \gamma$, $\pi e \nu$,
and $3\pi$
final states.
We used its measurements to
reject
background events in the $\pi \pi$
sample from semileptonic and $3\pi$ decays, and from $K_S$
produced through processes other than coherent regeneration.
$K$ mesons produced in a beryllium target traveled through
a series of collimators and sweeping magnets before reaching the regenerators.
Data from a
drift chamber spectrometer and lead glass electromagnetic calorimeter
provided kinematic information about charged particles
and photons in the final state.  Signals from scintillation counters
contributed to the experiment's triggers and vetoes.
A diagram of the detector is shown in Fig.~\ref{e773detector}; individual
subsystems will be described below. The thin scintillator planes
labeled ``T,V" inside the vacuum
were removed midway through the run. The transition radiation detector (TRD)
downstream of the last drift chamber was not used in the analysis of E773
data.

\subsection{Neutral beam}
\label{sec:beam}
Fermilab delivered a spill of 800~GeV protons to the E773 target about once per
minute. The protons arrived in 2~ns-long ``buckets" spaced by 19ns; each
spill lasted twenty seconds and contained typically $1.6 \times 10^{12}$
protons.  A neutral beam was formed by placing sweeping magnets, followed by
a 5.8m-long copper collimator, downstream of the 36~cm long target.
The collimator contained two tapered channels which created a clear path for
particles traveling 4.8 milliradians to the east of, and slightly above
or below, the incident proton beam's
direction.
Beryllium blocks placed downstream of the two-hole collimator,
together with
the non-zero targeting angle, reduced
the beam's neutron content relative to its kaon flux.
Some of the beryllium blocks shadowed only the downstream regenerator,
reducing the hadron flux striking it relative to that in the beam containing
the upstream regenerator.
A 7.6~cm thick lead block (13.6 radiation lengths)
converted photons to electron-positron
pairs which were removed from the beam by subsequent collimators and sweeping
magnets.
With the exception of gaps in the vacuum system at the
regenerators,
the beams traveled in vacuum from 17~m downstream of the target until
arriving at the first drift chamber, 159~m from the target.
By the time the beams reached the lead glass array, 181~m from
the target, each beam was a square, approximately
8~cm on a side.  The beams
were separated vertically with
$\sim$15~cm clearance between their
inner edges, and passed through holes in the lead glass array, stopping in
the 3.2~m steel muon filter.
Fig.~\ref{kebeams} shows the kaon beam profiles at the lead glass,
determined from $K_L \rightarrow \pi^+\pi^-$ decays.
(As with most of the figures, the histogram represents
data while the superposed points indicate the predictions of the
Monte Carlo simulation.)
Approximately $10^7$ $K_L$'s (and an equal number of neutrons) struck
the regenerators each spill.
The $K_L$ energy spectrum is shown in Fig.~\ref{k_energy_spectrum}; the
detector's acceptance allowed us to use
$K$ mesons with energies above $\sim 25$~GeV and below $\sim 160$~GeV.

\subsection{Regenerators}
\label{sec:regenerators}
In order to allow efficient rejection of events in which a $K_S$
or $K_L$ was produced
through inelastic processes, both regenerators were built
entirely from small
blocks of plastic scintillator. The upstream regenerator
was 118~cm long with one quarter of its blocks viewed by photomultiplier
tubes. The downstream regenerator was 40~cm long with
each of its blocks viewed
by a pair of phototubes.  Signals from some of the phototubes participated
in the experiment trigger; all signals were measured by ADC's and recorded
with event data. Straight-through muons were used to cross-calibrate
the regenerator channels.
The moving machines which positioned the regenerators (and
the beryllium absorber which shadowed the downstream regenerator)
were controlled by one
of
the data acquisition computers.
The regeneration amplitudes for 80~GeV kaons
were $\rho_{up} \approx 0.02e^{-i31^\circ}$
and $\rho_{down} \approx 0.007e^{-i33^\circ}$
in the upstream and downstream regenerators;
the magnitude of $\rho$ decreased
with increasing kaon momentum as $\sim p^{-0.6}$.

\subsection{Tracking system}
\label{sec:tracking}
The drift chamber spectrometer consisted of two pairs of wire
chambers\cite{woods}
which measured particle trajectories upstream and downstream of a large
analyzing magnet.
The chambers' cells were hexagonal, with a maximum drift distance of 6.35~mm.
Half-cell offsets between paired sense planes provided
resolution of ``left-right" ambiguities.
The chambers used an argon-ethane gas mixture which included a small amount
of alcohol vapor. Single-hit TDC's with 1~ns resolution digitized
discriminated chamber signals in response to an experiment trigger.
Chamber resolution was better than 100~$\mu m$
per sense plane; frequent magnet-off runs of
muon data were used to generate alignment constants for the tracking system.
Helium bags placed between the chambers served to
reduce the effects of multiple scattering.
Average chamber efficiency was typically 99\% per sense plane, but
a small number of wires showed lower efficiencies. The
effects of inefficient wires were modeled in the Monte Carlo simulation
of the detector.

The analyzing magnet's vertical field
imparted a transverse momentum kick of 200~MeV/c.
Its field integral had been
mapped on a two inch grid for E731; the magnet was run at the
same field strength in E773. The magnet's aperture was 1.46~m high by
2.5~m wide by 1~m deep and contained a helium bag during running.

The tracking system's momentum resolution was typically better than
1\%, with a momentum dependence given by
$${{\sigma_p} \over p} = \sqrt{(0.45\%)^2 + (0.012p\%)^2}$$
for the momentum $p$ measured in GeV/c.
The energy scale was verified by comparing the reconstructed masses in
$K \rightarrow \pi^+ \pi^-$ and $\Lambda \rightarrow p \pi$ decays with
the known $K$ and $\Lambda$ masses.

\subsection{Lead glass}
\label{sec:lead_glass}
%
The energies and positions of electrons and photons were measured by
a lead glass calorimeter built from 804 blocks of Schott F-2 glass.
Individual blocks were 5.8~cm square by 61~cm deep and
were stacked
parallel to the beam direction in a nearly circular array.
The radius of the calorimeter was about
0.92m; the beams passed through a pair of
holes near the center of the array.
The array was 18.74 radiation lengths deep.
Over the course of the run, blocks near the beam holes became less transparent
because of radiation damage.  Periodically, it was necessary to ``cure"
the blocks with exposure to ultraviolet light
from a mercury vapor lamp which resulted
in significant (but not complete) recovery.  The entire calorimeter was
located in a temperature controlled, light-tight enclosure.

An Amperex 2202 photomultiplier tube was mounted on the downstream
face of each lead glass block to collect \v Cerenkov light from
electromagnetic showers. Wratten 2A filters between
blocks and phototubes transmitted
light with wavelengths greater than 430~nm. (For
shorter wavelengths, the absorption in the lead glass was substantial
and
varied rapidly with wavelength.)  Optical fibers, mounted on the
upstream faces of all blocks, carried light from a single xenon
flash lamp which allowed us
to monitor detector performance.
All blocks were constantly illuminated
by a small
amount of light from a light-emitting diode
which served to
minimize the rate dependence of phototube gains.

Signals from the lead glass
photomultiplier tubes were integrated for 150~ns and digitized by
LeCroy 1885 ADC's. These dual-range ADC's had conversion gains that
were typically 5~MeV per ADC count below 17~GeV,
and 40~MeV
per count above
17~GeV.
Information from channels with $\geq$5 counts
was recorded with event data.

Since errors in the measurement of photon energies
induced errors in the determination of $\pi^0\pi^0$ decay vertex
positions, it was necessary to understand
in detail the calorimeter's behavior.
We calibrated the lead glass array with
20 million electrons from the decay $K \rightarrow \pi e \nu$ ($K_{e3}$).
A clean
electron sample was obtained after eliminating possible backgrounds
from
$\Lambda \rightarrow p \pi$,
$K \rightarrow \pi^+ \pi^-$, and $K \rightarrow \pi^+ \pi^- \pi^0$ decays
with kinematic cuts.
$K \rightarrow \pi \mu \nu$ ($K_{\mu 3}$)
decays were discarded by requiring that no signals
be present in the muon veto system.
Additional cuts required that tracks were well-reconstructed and
well-separated at the lead glass. Cuts on transverse shower
shape and the
maximum distance between the cluster position and the projected
electron track helped eliminate events
in which electrons lost energy through bremsstrahlung
emission.

The reconstruction of electromagnetic
showers depended on three parameters which had
to be determined for each block: two effective gains and
a parameter
quantifying a block's opacity to \v Cerenkov light.
An iterative procedure was used to
determine calibration constants for
each block;
the procedure looped repeatedly over
electrons in a calibration sample, updating constants after each pass,
until parameter values stabilized.
The lead glass calibration
parameters changed with time during the run.
Central blocks
experienced the most radiation damage and needed to have their
``constants" updated most frequently. Fortunately,
$K_{e3}$ electrons hit blocks near the center of the array most
frequently, allowing us to track
the most rapidly changing parameters.  We were able to
generate new calibration constants for central blocks
from approximately one day's running.

Fig.~\ref{eoverp}
shows the final ratio of the calorimeter energy
and track momentum (E/p)
for $K_{e3}$ electrons. Superposed on the figure is the
Monte Carlo simulation's $E/p$ spectrum
(see Sec.~\ref{sec:montecarlo});
the r.m.s. width of the peak is approximately 3\%.
The distributions agree except for
the low side tail in the data which is not yet well-represented by the Monte
Carlo.
The stability of our calibration
over time is
shown in Fig.~\ref{eoverpvst}.
The
period of time spanned by
the plot represents about nine weeks, the length of
E773's data run.
Fig.~\ref{eoverpvsp} shows the results of fits for the
mean and width of the $E/p$
distribution as a function of track momentum.
Events were restricted to the region near the
peak;
the apparent shift of the mean away from unity is an
artifact caused by this
restriction.

\subsection{Trigger and vetoes}
\label{sec:trigger}
The principal E773 triggers and vetoes were designed to select events
with two tracks (``charged mode") or four photons (``neutral mode")
in the final state. Additional triggers,
useful for calibration and monitoring purposes,
selected single muon events, six photon
events, and ``accidentals," events which gave us an indication of the level
of additional activity in the detector which might be expected
to accompany signals produced by the
decay products of real $K$ mesons.

Some parts of
the veto system participated actively in triggering decisions for both charged
and neutral mode triggers.  For example, signals in
the downstream scintillators in either regenerator (which were expected
to be quiet during coherent regeneration) would block both modes' triggers.
Other parts of the veto system
rejected triggers caused by
random detector activity
as well as triggers
from
events such as $K_{\mu 3}$ decays which were accompanied by muons.
(We used electrons from the copious decay
$K_L \rightarrow \pi e \nu$ to calibrate the
lead glass and to verify the accuracy of the $\pi^+\pi^-$ Monte Carlo
simulation.)  Vetoes for the neutral trigger allowed us to reject
events
in which
energy depositions
in the lead glass came from hadrons striking the calorimeter. Other neutral
mode vetoes permitted us to discard triggers in
which significant electromagnetic
energy went down the beam holes in the lead glass array.
Analogue signals from most veto
counters were measured by ADC's and recorded with event data.
We used this information during offline analysis to
discard events
in which photons might have missed the lead glass array, based on
activity in the photon veto system.
To reduce backgrounds from inelastic $K_S$ production,
neutral and charged mode events were discarded when accompanied by
activity in either regenerator.

The charged mode used a two-level trigger.
In the first half of the run, the level 1
trigger required a signal in any counter in the T,V scintillator
hodoscope in coincidence with at least
two signals
from scintillators in the B hodoscope and at least two from counters
in the C hodoscope
(See Fig.~\ref{e773detector}).  The pattern of hits in the B and C
counter banks
was required to be consistent with a pair of tracks from a two-body decay in
one of the beams.
A signal in the $\mu 2$ hodoscope (located downstream of a 3.2~m-thick steel
wall) would veto a first level trigger, allowing us to reject
$K \rightarrow \pi \mu \nu$ decays.  The presence
of signals in photon veto counters immediately before the lead glass
would also block the trigger, allowing us
to discard two-track events which might have come from
$K \rightarrow \pi^+ \pi^- \pi^0$ decays.

The level 2
charged mode trigger in the first half of the run
was based on the pattern of struck wires in the drift chambers.
The trigger required that
the four chambers each
have at least one
hit in
both left and right sides, as expected in a two-track final state.
Midway through the run, the T,V hodoscope was removed and a
more sophisticated second level trigger installed.  The new trigger
required that the pattern of hits in all four chambers was consistent with
that expected from a two-body, two-track event, with the possible inclusion
of extra hits from accidental
activity. Both $\pi^+ \pi^-$ and $\pi^+ \pi^- \gamma$ final states satisfied
the
charged trigger; no requirement was made on lead glass energy.

The neutral mode also used a two-level trigger.
The first level required that the total energy in the lead glass
be above $\sim 25$~GeV.
Outputs from groups of nine contiguous lead glass blocks were
summed by ``adders;" individual adder outputs were integrated
for 30~ns and digitized by ADC's.
(This shorter gate helped us eliminate
``out of time" clusters registered by the lead glass ADC's.)
These sums were combined to form the total energy,
$E_T$, as
a fast analogue sum of signals from the entire lead glass array.
The $E_T > 25$~GeV requirement was effective since most
accidental activity in the lead glass
came from muons or out-of-time showers,
which tended to deposit small amounts of
energy in the calorimeter.

The second level neutral trigger employed a ``hardware cluster finder"
(HCF) to count the number of isolated depositions
of energy (clusters) in the lead
glass.\cite{hcf}
Events with four clusters were accepted
as candidate $\pi^0 \pi^0$ events.

The level 1 neutral mode vetoes
were intended to reject triggers associated with hadronic activity
in the lead glass or with electromagnetic activity in the
vicinity of the calorimeter's beam holes.
A twenty-one radiation length lead wall behind the lead glass
prevented
electromagnetic showers in the calorimeter from registering in the
$\mu 1$ hodoscope.
Since hadronic showers in the lead glass (which could fake a
four-photon trigger) would usually produce activity in
$\mu 1$, a large signal in $\mu 1$
served to veto a level 1 neutral trigger.

A twenty-eight
radiation length lead-lucite shower counter, the BA (``back anti"),
was installed in the beams behind the calorimeter.
Electromagnetic showers tended to deposit
most of their energy in the upstream part of the
BA, while interactions
of beam neutrons and kaons
produced large signals towards the back of the BA.
Events with substantial activity in the upstream part of the BA, but minimal
energy in the last third of the BA, were vetoed.  The comparison of
the signals in the up- and downstream sections allowed us to
reject events with lost photons
without vetoing good events with accidental activity associated with
the interaction of a beam hadron in the BA.  Muons and $K_{e3}$
electrons provided calibration information about the BA.

The CA (``collar anti'') covered the inner halves of
the blocks around the two beam holes and consisted of eight
radiation lengths of material
followed by scintillation counters.  The CA
allowed us to veto an event in which
a photon would have landed near a beam hole;
the  reconstruction of the energy and
position of these photons would have been problematic.

Most of the photon veto counters were constructed of a plane of scintillator
followed by a
lead-lucite shower counter six
radiation lengths thick.
Each counter's scintillator was viewed by one phototube, while the shower
counter was viewed by a pair of phototubes.
These
counters
were designed to detect low energy photons which were
traveling at angles up to 50 milliradians with respect to the beams;
they were typically 90\%
efficient at registering 150~MeV photons.
We mapped the position dependence of the response of
photon veto counters with muons.  We were able to calibrate
their overall response using $K \rightarrow \pi^+ \pi^- \pi^0$
decays in which one photon missed the lead glass.\cite{magnus}
The direction and energy of the missing photon
could be inferred from the kinematics of the charged pions and reconstructed
photon.  Often this missing photon passed through one of the photon veto
counters.  By comparing the observed signals in the veto counters to the
predicted energy of the ``undetected'' photon, the gains and resolutions
of the
counters were extracted.
We included many of the photon veto scintillation counters in the
level 1 vetoes for charged and neutral triggers. However, only the
shower counters in the plane closest to the lead glass (``LGA")
participated in the level 1 vetoes. Information from the shower counters
was used during offline analysis to discard events in which photons
might have missed the lead glass.

In addition to the $\pi^+ \pi^-$ and $\pi^0 \pi^0$ triggers,
several special
purpose triggers were collected at the same time.
The ``accidental" trigger was used to study the effects of
activity in the detector not associated with a decaying $K$ meson.
Muons traveling through a scintillator telescope aimed at the target pile would
assert the trigger.  The telescope was roughly 50~m from
the target; the line from the target through the telescope passed
approximately ten meters west of the spectrometer.
As a result, the accidental trigger rate
was proportional to the instantaneous beam intensity, but
was otherwise
independent of the presence or absence of activity in the spectrometer.
Pedestal triggers were used to record the
zero-signal behavior of all ADC's; calibration flasher triggers were used to
monitor the behavior of the lead glass calorimeter.
Single muon triggers written during special
calibration runs
were used to monitor the performance of many
subsystems in the detector.

\subsection{Data acquisition}
\label{sec:DA}
A large number of signals were registered
by the detector's data acquisition electronics
for each event which satisfied the experiment trigger.
Discriminated signals from scintillation counters were stored in CAMAC
latches, while analogue information from the lead glass,
regenerators, and
veto system was processed by FASTBUS ADC's. All data, including drift chamber
hit times and status information from the trigger electronics, flowed into
a PANDA\cite{panda} data acquisition system which
could record approximately 13,000 events per
spill. PANDA wrote data simultaneously to four 8~mm cassette tapes and
passed a fraction of the events to a microVaX which
monitored the experiment and generated histograms.
Data describing magnet currents and targeting
information were stored at the end of each spill along with integrated
counting rates from various detectors and
components of the trigger.

\section{Reconstruction of $\pi^+\pi^-$ and $\pi^+\pi^-\gamma$ Events}
\label{sec:chgreconstruction}
\subsection{Tracking}
\label{sec:trackrecon}
Tracks in the horizontal ($x$) and vertical ($y$)
views were found independently.
Since the analyzing magnet deflected charged particles in the horizontal
plane, $x$-view track segments were
found separately in the pairs of drift chambers
upstream and downstream of the magnet. Tracks in the $y$-view were
found in all four chambers simultaneously.  Trial
track segments in the upstream chambers were required to point in the
general direction
of the decay volume, while downstream segments
were required to point towards the lead glass.
Hits were required
in at least three of the four sense planes in upstream and downstream
$x$-view track segments.
A $y$-view track was required to have hits
in at least five of eight sense planes.
No tracks were allowed to share hits.
After track segments were identified,
drift distance information from the TDC's was
used to refine the tracks' positions.

Track segments in the $x$-view were
projected to the ``bend plane" at the center of the analyzing magnet.
Upstream and downstream segments were paired if their projections
were separated by less
than 1~cm.
The $x$-view and $y$-view tracks were
correlated using
information from cluster positions measured in the lead glass.
After matching of horizontal and vertical views,
hit positions were recalculated to correct for small chamber
rotations,
gravitationally-induced sagging of sense wires,
and differences in signal propagation time along the sense wires.
At this point, upstream and downstream $y$-view track
segments were refit separately since the analysis magnet tended to
impart a
small vertical impulse to tracks. The difference in
upstream and downstream $x$-view track slopes, combined with knowledge of the
analyzing magnet's field integral, provided us with sufficient information
to calculate a
particle's momentum.

Since the paired sense planes in the drift chambers were
staggered by a half-cell, the sum of the measured drift distances in one
view of a chamber should equal the 6.35~mm separation between adjacent
sense wires. (Because of the 1.1~cm separation between sense planes along the
beam direction, a small correction was applied for track angle with
respect to the $z$ axis.)
The sums-of-distances allowed us to recognize
signals which came from out-of-time tracks. Given the temporal structure
of the extracted Fermilab proton beam (see Sec.~\ref{sec:beam}), a track from
an earlier (or later) bucket would have its sums-of-distances mismeasured by at
least 1.8~mm.

We required pairs of tracks to come from a common vertex.
The vertex position was determined using a full fit which included
individual tracks' estimated accuracy in contributing to the
vertex determination.
Our requirement on the tracks' distance of closest approach
depended on track momenta and on the distance from the
vertex to the chamber system.
Typically, tracks with a few millimeter
separation satisfied the vertex-quality criterion.
Vertex resolution was $\sim$0.8~mm in the $x$ and $y$
views, and $\sim15$~cm in $z$.
The vertex position in $\pi \pi \gamma$ decays was set equal to the
tracking vertex, without reference to the reconstructed photon.

Additional tracking cuts were made to ensure reconstruction quality.
A $\chi^2$ cut  rejected tracks with large deviations of hits
from ideal positions.  A matching ambiguity cut discarded events
in which the lead glass information could not be used reliably to
pair tracks with similar $x$ positions with the corresponding
$y$-view tracks. We also required
reasonable agreement in the projections
to the center of the magnet for up- and downstream track segments in
both $x$ and $y$ views.

\subsection{Event reconstruction}
\label{sec:pipirecon}
Events which came from
$K \rightarrow \pi^+ \pi^-$ and $K \rightarrow \pi^+ \pi^-\gamma$
decays were expected to reconstruct with
invariant mass close to the known kaon mass. Since we were interested
in detecting kaons which did not scatter in the regenerators,
we required that the final state $\pi^+ \pi^-$ or $\pi^+ \pi^-\gamma$
have minimal transverse momentum ($p_T$) with respect to
the incident kaon direction.
Almost all $K_{\mu 3}$ decays were removed at the
trigger level so most charged mode triggers came from
$K \rightarrow \pi e \nu$ decays.
There were also significant contributions to the event sample from
$K \rightarrow \pi^+\pi^-\pi^0$ and $\Lambda \rightarrow p\pi^-$ decays.
Events in which a kaon scattered in the regenerator
contributed to observed backgrounds.

Backgrounds from $K_{e3}$ decays were suppressed by rejecting events
which had a ratio of calorimeter energy and track momentum ($E/p$) close to
unity.
Requirements on $E/p$ for the $\pi^+ \pi^-$ and
$\pi^+ \pi^-\gamma$ analyses are described in Table~\ref{cuts}.
At the expense of losing a few
percent of the events with two charged pions in the
final state we were able to reject nearly all the $K_{e3}$ background.
We required tracks to pass through regions of the lead
glass in which cluster energies could be reconstructed reliably.
Events with
a track near the outer edges of the glass, in a beam hole, or in the
collar anti were discarded.

To reduce background contamination from $K_{\mu 3}$ decays,
we required both tracks to project into the $\mu2$ hodoscope.
Since ``soft" muons
were likely to stop in the 3.2~m steel muon filter rather than
reaching $\mu2$, we rejected events in which either track
had momentum below 7~GeV/c.
This cut also reduced the
fraction of $\pi^+ \pi^-$ decays which were
vetoed because a pion decayed in flight
with the resulting muon striking a $\mu2$ counter.

Only the
most energetic $\Lambda$'s from our target
survived long enough to decay downstream of the
first regenerator. Most of these interacted
in a regenerator or were eliminated by the requirement that
tracks neither hit the collar anti nor went
through a beam hole in the lead glass.
The majority of $\Lambda \rightarrow p \pi$ decays
came from hadronic interactions in the regenerators
which produced a $\Lambda$ without asserting the regenerator
vetoes.
Because of the small Q value, the ratio of the proton and
pion momenta in high energy $\Lambda$ decay will
always satisfy $p_p/p_\pi > 3.09$.
Since we had no particle
identification for separating hadronic species, we assumed that the
track with greater momentum was a proton (or antiproton)
when reconstructing a two-track state as a $\Lambda$ (or $\bar{\Lambda}$).
Kinematic cuts (to be described below) were used to reject $\Lambda$
decays.

To reduce sensitivity to imperfections
in our simulation of accidental activity in the
detector,
both the $\pi^+ \pi^-$ and $\pi^+ \pi^- \gamma$ analyses
required that reconstructed tracks
satisfied the charged mode trigger.
In addition to this trigger ``verification," we
discarded events where
tracks passed near
limiting
apertures.

Values for various cuts in the $\pi^+ \pi^-$ and $\pi^+ \pi^- \gamma$
analyses are listed in Table~\ref{cuts}.

\subsection{$\pi^+ \pi^-$ signal and backgrounds}
\label{sec:pipisig}
Reconstructed $K \rightarrow \pi^+\pi^-$ mass
distributions (after all other cuts,
including $p_T^2$)
are shown in Fig.~\ref{pipimass}.
Events from the upstream and downstream regenerator
beams are plotted separately.  In addition,
data from
the periods before (set 1) and after (set 2) the removal of the
T,V hodoscope are plotted separately.
Results from the Monte Carlo are superposed as points
on the data histograms.
The data's high-side tails contained
contributions from $\delta$-rays which were not simulated by the Monte
Carlo.
Our mass resolution was $\alt 3.2~{\rm MeV/c^2}$ in data
written before the T,V hodoscope was
removed and $\sim 0.25~{\rm MeV/c^2}$ better after its removal;
the analysis required
$484\ MeV/c^2 < m_{\pi \pi} < 512\ MeV/c^2$.
The tail on the
low side of the peak came from radiative $K\rightarrow\pi\pi\gamma$
decays.
Backgrounds from
semileptonic
decays were nearly negligible.
There was no background contribution from
$\pi^+\pi^-\pi^0$ decays since the invariant mass
of the
$\pi^+\pi^-$ system was well below the signal region.
Events which had
$1110\ {\rm MeV/c^2} \leq m_{p \pi} \leq 1122\ {\rm MeV/c^2}$
when reconstructed as $\Lambda \rightarrow p \pi$ candidates
were discarded.

We calculated the apparent transverse momentum of a
kaon with respect to the incident beam particle's direction
in order to reject backgrounds from diffractive and inelastic
$K$ production as well as semileptonic decays.
We extrapolated the
kaon trajectory from the decay vertex
upstream to the
regenerators.
The kaon's transverse momentum, $p_T$, is the
component of the observed vector momentum which is
perpendicular to the line connecting the production
target with the kaon's position at a regenerator.
Because of the spacing between the beams' inner edges
we could determine unambiguously which regenerator
the kaon traversed.

The $p_T^2$ distributions are shown in
Fig.~\ref{charpt}.  The $\pi\pi$ analysis required
$p_T^2 < 250~(MeV/c)^2$; tails near the coherent regeneration
peak in the distribution
were dominated by radiative
$K\rightarrow\pi\pi\gamma$ decays
and diffractively regenerated
$K_S$. At large $p_T^2$, inelastically produced $K_S$
contributed significantly to the tails.
Some smearing of the data's coherent peak was due to the
presence of $\delta$-rays.

The $p_{T}^{2}$ distributions of events
passing all other cuts were
used to determine background levels.
The small
backgrounds consisted of a mix
of diffractive $K \rightarrow \pi^+ \pi^-$ decays,
radiative $\pi^+ \pi^-\gamma$ decays, and
semileptonic final states; they are described in
Table~\ref{pipibkgrd}.
A total of 1.8 million $\pi^{+}\pi^{-}$ decays were found
in the entire run's data after all cuts;
the event sample is described in Table \ref{pipievents}.
We required that kaons had energies in the range 30 to 160~GeV
and that
decay vertices were downstream of the vacuum
window following the appropriate regenerator.
To reduce backgrounds associated
with interactions of beam hadrons in material near the downstream
regenerator,
we discarded events from the upstream regenerator beam
with decay vertices near the gap in the vacuum system.
Figs.~\ref{pipiz} and
\ref{pipie} show the decay vertex $z$ position and $K$ energy
spectrum for $\pi^+ \pi^-$ decays passing all other cuts.
Data (histogram) and Monte Carlo (points) are shown for each beam.
The T,V hodoscope, located near $z=141$~m,
was included in the charged mode trigger
for set 1 data.

\subsection{$\pi^+ \pi^- \gamma$ signal and backgrounds}
\label{sec:pipigsig}
The $\pi^+ \pi^-\gamma$ analysis required two well-reconstructed tracks and
at least one in-time lead glass cluster with energy
$\geq$1.5~GeV which was not associated with either track.
In the center of mass of the kaon, this photon was required
to have an energy greater than 20~MeV.
To suppress background from $\pi^{+}\pi^{-}\pi^0$
decays, we required
\[
\frac{\left[ \left( m_{K}^{2} - m_{0}^{2} - m_{+-}^{2} \right)^{2} -
4m_{0}^{2}m_{+-}^{2}  - 4m_{K}^{2}\cdot (p_{+-}^{T})^{2} \right]}
{4\left[(p_{+-}^{T})^{2} + m_{+-}^{2} \right]}\ \equiv
\ p_{0}^{2} \ <\ -0.01125\ .
\]
Here, $m_{+-}$ is the invariant mass of the two charged tracks
and $(p_{+-}^{T})^{2}$ is the square of their
transverse momentum with respect to the
parent kaon. (We assumed that the kaon traveled in a straight line
from the production target to the decay point and did not scatter
in the regenerator.)
Since
$p_{0}$ is proportional to the $\pi^0$ momentum component in the
kaon center-of-mass which is parallel to the beam direction,
$p_{0}^{2} \geq 0$ for events which might be $3\pi$
decays.~\cite{pp0kin,Carroll2} Shown in Fig.~\ref{estargamma}
are energy spectra for the photon in the $K$ center of mass
for $\pi^+ \pi^- \gamma$ decays satisfying all other cuts.

We removed background from
$K \rightarrow \pi^+ \pi^-$ events
accompanied by accidental activity in the calorimeter
by discarding
events with $m_{+-} > 484 {\rm MeV/c^2}$.
Remaining backgrounds from $\Lambda \rightarrow p \pi$ decays
(in combination with an accidental lead glass cluster)
were reduced with kinematic cuts.
Events which reconstructed to have
$1100\ {\rm MeV/c^2} \leq m_{p \pi} \leq 1130\ {\rm MeV/c^2}$,
total energy above 100~GeV, {\underline {and}}
momentum ratio $p_p/p_\pi > 3.0$
were discarded.

To select events in which only coherently
regenerated $K_S$ and $K_L$ decay amplitudes participated, we
rejected events with reconstructed mass outside the range
${\rm 484\ MeV/c^2} < m_{\pi^+\pi^-\gamma} < {\rm 512\ MeV/c^2}$
or with
large transverse momentum such that $p_T^2 > {\rm 150\ (MeV/c)^2}$.
Shown in Figs.~\ref{pipigmass} and ~\ref{pipigptsq}
are mass and $p_T^2$
for $\pi^+ \pi^- \gamma$ decays satisfying all other cuts.
Results of the Monte Carlo
simulation (shown as points)
did not include
background modeling.

The data separated naturally into two sets corresponding
to the periods of running time
before and after the removal of the T,V hodoscope.
After cuts and background subtractions,
a total of (10,769 $\pm$ 17) $\pi^{+}\pi^{-}\gamma$ decays were found
in the two data sets;
the distribution of events between the sets
is shown in Table~\ref{events}.

Events for each set were binned according to kaon momentum $p$
and decay position $z$ in a grid with cell size 10~GeV/c $\times$ 2~m.
We required kaons to have energies between 25 and 155~GeV.
Figs.~\ref{pipigz} and
\ref{pipige} show the decay vertex $z$ position and $K$ energy
spectrum for $\pi^+ \pi^-\gamma$ decays passing all other cuts.
Decay vertices were required to be downstream of the vacuum
window following the appropriate regenerator.
The mass and  $p_{T}^{2}$ distributions of events
passing all other cuts were
used to determine the level of background in each set.
Backgrounds were typically 2\%, and are listed
for each subset in Table~\ref{bkgrd}.  The small
backgrounds consisted of a mix
of noncoherent $K \rightarrow \pi^+ \pi^- \gamma$ decays and final states
made from two-track events combined with accidental activity in the
calorimeter. We modeled
the shape of the background
using events which passed the $\pi\pi\gamma$ mass cut, but
had
$p_{T}^{2}$ between 300 and 2000 (MeV/c)$^{2}$.
These events were binned in $(p,z)$ grids identical
to those of the signal region,
scaled appropriately, then
subtracted from the signal-region $(p,z)$ grids.

\section{Reconstruction of $\pi^0 \pi^0$ Events}
\label{sec:neureconstruction}
\subsection{Photon reconstruction}
\label{sec:clustering}
The reconstruction of an electromagnetic
shower (``cluster") in the lead glass array proceeded in stages.
Lead glass channels whose ADC signals were local
maxima were taken as the central
blocks in candidate clusters.  Signals from
central blocks were required to be above 200 ADC counts;
signals from
the eight blocks surrounding each central block
were included in the energy estimate formed for
clusters.

A first estimate of a cluster's energy was made using the
sum of the ADC counts,
divided by the
effective gains of each channel, in the cluster's nine blocks.
The effective gain
included photomultiplier gain, ADC digitization
conversion gains, photocathode quantum efficiency, and light
collection efficiency.
A rough correction for absorption of \v Cerenkov light in each block
was applied to the initial cluster energy estimate.
With this
correction, the estimated cluster energy
was usually within 30\% of the true shower energy.

Several corrections were made to this initial estimate.
A small amount of energy was added to account for blocks
which did not meet the readout threshold of five ADC counts.
A second correction compensated
for radial leakage outside the nine blocks in a cluster, into the beam
holes, or past the outer edges of the array.
When showers overlapped, the energy in shared blocks was divided
between showers based on typical shower profiles.
During the early part of data taking, the lead glass enclosure's
temperature control system malfunctioned.  Since we recorded
the enclosure's temperature with event data, we
were able to
correlate gain changes with temperature, and to map the
enclosure's temperature as a function of time.
Using this information, we
applied a small time-dependent
gain correction to lead glass signals.
The ADC pedestals showed a small rate-dependent
shift; we corrected them
based on the instantaneous event rate.

The final determination of the energy in
an electromagnetic
shower
included correction for effects associated with absorption
of \v Cerenkov light in the lead glass.
The depth at which a shower deposited most of its light
increased with shower energy. As a result, the response of the
calorimeter tended to ``brighten" with increasing shower energy, since
less light was absorbed by the glass between ``shower max" and
the photomultiplier tube.
The absorption coefficients varied from block to block, and changed with
time as
radiation damage increased. The coefficient of the cluster's central
block was used to calculate the correction factor.
An additional correction compensated for the fact that
photons began showering deeper in the lead glass than did
electrons.
Shower energies associated with photons which converted to $e^+e^-$
pairs in the B and C hodoscopes were adjusted by 3.5\%.
In a final correction we shifted
electron-induced shower energies 0.3\%
to compensate for the effects of
bremsstrahlung photons produced in the trigger
hodoscopes near the calorimeter.

Shower $x,y$ positions were found
using the estimated energies deposited in the three columns
and three rows
of blocks in a cluster.
We studied the calorimeter's position resolution by
comparing the impact point of electrons, determined by
the drift chambers, with shower centers, determined by the lead glass.
The position algorithm's results were typically accurate
to 3~mm, though the resolution varied from 1.5~mm near the corners of
blocks to 5~mm at the centers of blocks.

\subsection{Event reconstruction}
\label{sec:pi0pi0recon}
The invariant mass of two photons originating from a single decay
vertex and traveling
at small angles relative to the beam direction is
$$m={r_{12} \over z} \sqrt{E_1 E_2}$$
where z is the distance from the decay vertex to the calorimeter,
$E_1$ and $E_2$ are the photons' energies,
and $r_{12}$ is their separation
at the calorimeter.  We were interested
in reconstructing final states in which two (or three) $\pi^0$'s
produced four (or six) detected photons.
There were three possible ways to ``pair" the photons in a candidate
$\pi^0\pi^0$ event, and fifteen ways in a candidate $3\pi^0$ event.
For each of the
possible pairings, we calculated a separate decay $z$ for each
pion by assuming that
the invariant mass of the corresponding
photon pair equaled the $\pi^0$ rest mass.
We then determined errors for each $z$
from
the uncertainties in cluster energies and positions, and calculated a
$\chi^2$
for the hypothesis that the photon
pairs came from the same vertex.
Using the pairing with the smallest $\chi^2$, we
assigned a decay $z$ for the event based on a weighted
average of the pions' individual $z$ positions.
Typical vertex resolution was about 1 meter;
the invariant mass of the four (or six) photons was
\begin{equation}
m = {1 \over z}\left({{\sum_{i<j} E_i E_j r_{ij}^2} } \right)^{1 \over 2}
\label{pairingmass}\ .
\end{equation}

Several cuts were made to reduce background contamination in the
data.
We rejected events with pairing $\chi^2 > 4$.
In addition,
we discarded four-photon events
in which the
second-best pairing had a reasonable $\chi^2$
and
yielded an
invariant mass near the $K$ mass.
Background from $3\pi^0$ decays with undetected photons
was reduced by discarding events with significant activity in
the photon veto system.
In addition, $3\pi^0$ decays with photon energies below the HCF thresholds
were removed by cutting events
with
$\geq 600$~MeV of in-time energy in any lead glass block
which was
not associated with one of the four reconstructed
clusters.
Backgrounds associated with interactions of beam hadrons with
material in the detector
were
reduced by cuts on the number of drift chamber hits
and on trigger hodoscope activity. As was the case with charged mode events,
we rejected
$\pi^0 \pi^0$ candidates with significant activity in either regenerator.

Other cuts were made
to simplify Monte Carlo simulation of the detector.
Photons were required to have energies greater than
2.2~GeV to reduce sensitivity to inaccuracies in the
simulation of HCF thresholds.
Events were discarded if any cluster reconstructed within
a half block of the outer edge of the lead glass array.
We rejected events with appreciable activity in the collar anti.
To avoid problems from events with overlapping photons,
we required the energy distribution among the blocks in a cluster
to be consistent with the shape expected for a single electromagnetic
shower.

The calorimeter was calibrated with electrons, not photons. The
coupling between the energy and $z$ scales in the invariant mass
relationship [Eq.~(\ref{pairingmass})]
was used to check the photon energy scale. The
reconstructed positions of the downstream faces of the regenerators
were studied as
functions of energy; the energy scale was adjusted
so that the data and Monte Carlo distributions
agreed.

Cuts made in the $\pi^0 \pi^0$ and $\Delta \phi$ analyses are summarized
in Table~\ref{deltaphicuts}.

\subsection{$\pi^0 \pi^0$ signal and backgrounds}
\label{sec:pi0pi0sig}
Our preliminary result for
$Arg(\eta_{00}) - Arg(\eta_{+-})$
is based on data written after the T,V hodoscope was removed.
(See Sec.~\ref{sec:trigger}.)
The reconstructed $K \rightarrow \pi^0\pi^0$
mass distributions (after all other
cuts) are shown in
Fig.~\ref{2pi0mass}.
Results from the Monte Carlo simulation of signal and backgrounds are
superposed on the data.
In the plots
on the left, the simulation includes both
signal and backgrounds. In the plots on the right, only the
simulated backgrounds are superposed on the data.
The signal region is between 474~MeV and 522~Mev. The
background is mostly from $K \rightarrow 3\pi^0$ with two undetected
photons. There
is a small background contribution
from neutron interactions in detector
material which produced two neutral pions.

The $x,y$ vertex position was not measured
in neutral
decays so the transverse momentum of the parent kaon, $p_T$,
could not
be determined directly.  Instead, the center-of-energy of the four
photons was used to select
decays of kaons which had not scattered in the
regenerators.
The center of energy was defined as
\[
\vec r_{C.E.} \equiv {{\sum E_i \vec r_i} \over {\sum E_i}}\
\]
where $\vec r_i$ was the point at which the $i^{th}$ photon struck
the calorimeter.
Since $\vec r_{C.E.}$ corresponded to the point at which the
kaon would have reached the lead glass if it had not decayed,
unscattered kaons had $\vec r_{C.E.}$ inside the beam spots.

We defined an event's ``ring number" to be the area circumscribed
by a square,
centered on the closer beam, whose perimeter contained $\vec r_{C.E.}$ (see
Fig.~\ref{ringdiagram}).
The ring number distributions (after all other cuts) are shown in
Fig.~\ref{ringplot}. Unlike the charged mode $p_T^2$
spectra, the neutral mode ring plots contain no sharp peaks because
of the finite beam size. Events were required to have a ring number
less than 120~${\rm cm^2}$. Near the coherent regeneration region in the
ring plot is
a shoulder from diffractively regenerated $K_S$; inelastically
regenerated $K_S$ populate the high ring number region.
As before, the predictions of the Monte Carlo
are
superposed on the data.

Normalizations for our background subtractions were
established using events which lay outside the signal
region in the mass $vs.$ ring number plane.
The background shape in the signal region was known from
Monte Carlo studies of $K \rightarrow 3 \pi^0$ decays which
faked $\pi^0 \pi^0$ events and from measurements of the $p_T^2$
distribution for $\pi^+ \pi^-$ events. (Our $p_T^2$
resolution in the charged decay mode
let us identify scattered kaons whose centers-of-energy
remained inside the beam profile.)
A total of $2.6 \times 10^5$
$\pi^0\pi^0$ decays were found in the set 2
data
after all cuts. The
event sample is described in Table~\ref{p0p0events}.
The kaon energy was required
to be in the range $40 \leq E_K \leq 150$
and the decay vertex z position
was required to be in the range $120 \leq z \leq 152$ for the
upstream regenerator and $130 \leq z \leq 152$ for the
downstream regenerator.
The vertex position distributions and
kaon energy spectra for $\pi^0\pi^0$ decays passing all other
cuts are shown in Figs.~\ref{p0p0z} and \ref{p0p0e}.

\section{Monte Carlo Simulation}
\label{sec:montecarlo}
We measured the phases of $\eta_{+-}$, $\eta_{+-\gamma}$,
and $\eta_{00}$ by studying the proper time evolution of
the $\pi \pi$ and $\pi\pi\gamma$ decay rates
downstream of the regenerators.
The detector's acceptance varied with kaon energy and decay
position, so we needed to unfold apparatus effects from observed
decay spectra in order to determine values for the
magnitudes and phases of the $\eta$'s.
We used a detailed Monte Carlo simulation
of the experiment to calculate the $p$ and $z$ dependence of the
detection efficiency.
We also used the Monte Carlo to study backgrounds and to determine
the effects of various sources of systematic uncertainty.
The accuracy of the simulation could be
verified with the large
samples of $K_{e3}$, $\pi^+\pi^-\pi^0$, and $3\pi^0$ decays
which satisfied the experiment triggers.

The Monte Carlo program generated raw data which
was processed by the same reconstruction programs
used to analyze real data.  The effects of spurious detector
activity could be studied by combining data from accidental
triggers (see Sec. \ref{sec:trigger}) with simulated
event data.

\subsection{Beamline simulation}
\label{sec:neutral_beam_MC}
The neutral beam simulation
modeled the creation of $K^0$ and ${\overline {K^0}}$
in our production target as well as the propagation of kaons through
absorbers and
the (imperfectly aligned) collimation system.
Even the small amount of residual $K_S$
amplitude present in high energy kaons striking our regenerators
was modeled by the beam simulation.
Propagation of kaons in the regenerators was
treated in detail, allowing for coherent regeneration, diffraction
regeneration through single (and multiple) elastic scattering,
inelastic processes,
and decays inside the regenerator.
Regeneration in air gaps, vacuum windows,
and the T,V hodoscope was included in
the Monte Carlo.
The simulation
of the interactions
of daughter particles from $K$ decay
included multiple Coulomb scattering, the production of
bremsstrahlung photons by electrons and positrons, and
the conversion of photons to $e^+ e^-$ pairs.
Simulated charged pions were allowed to
decay in flight while traveling
through the spectrometer.

We based our $K^0$, ${\overline {K^0}}$
energy spectra on Malensek's parametrization of the
$K^+$ and $K^-$ production spectrum for 400~GeV protons incident
on a beryllium target\cite{malensek}.
He described the differential production cross section in
terms of $x_F \equiv p_K / E_{beam}$ and $p_T$; in general, cross sections
expressed in terms of these variables scale well with beam energy
(Ref.~\cite{PDG}).
Because of measurement errors in the result of Ref.~\cite{malensek}
and uncertainties associated with the extrapolation to higher incident
beam energy, we found it necessary to adjust the
shape of the $K$ production spectrum to fit our data.  The
correction function varied from unity
by $\sim \pm 20\%$ over the range of usable
kaon energies.
Though it was very convenient to have the Monte Carlo's energy spectrum
well-matched to the data's energy spectrum,
our analysis was not particularly sensitive to
inaccuracies in spectrum modeling.

The Monte Carlo propagated kaons through the target and the beryllium
and lead absorbers (see Sec.~\ref{sec:beam}), modeling
the effects of coherent regeneration
and (non-regenerative) single elastic scattering
in addition to attenuation of the beam.
The simulation modeled the
$x,y$ intensity profile of the incident proton beam
and allowed kaons to
scatter in collimator walls.
The accuracy of the absorber/collimator description can
be seen in Fig.~\ref{elliottbeams} for $K_{e3}$ decays. Tails
in the data's
distribution of
angle between the target-to-vertex direction
and the
$z$ axis
are well-matched by the Monte Carlo.
These tails are associated with elastic scattering
of kaons in beamline elements.

The $x,y$ locations and sizes of
limiting apertures in the detector were measured using electrons from
$K_{e3}$ decays,
while
the $z$ positions of the apertures
were determined using survey information.
Shown in Fig.~\ref{elliotthdra} are comparisons of the
data and Monte Carlo electron illuminations at the east edge of
the HDRA aperture in 1~mm bins.  Illumination of charged pions
at the HDRA and lead glass are shown in Figs.~\ref{chgillumhdra} and
\ref{chgillumglass} for events which passed $\pi^+ \pi^-$ analysis cuts.

\subsection{Drift chamber simulation}
\label{sec:tracking_MC}
The Monte Carlo simulation of the drift
chambers did not attempt to model detailed processes
like  electron drift and diffusion.
Instead,
the drift time distribution observed in the data
was integrated to produce a mapping of
drift distance $\leftrightarrow$ drift time.
Chamber resolution was incorporated by jittering the true
positions of particles at chamber sense planes with the
appropriate Gaussian
distributions.
Smeared positions were then converted to drift times.
When two tracks passed through the same drift cell, only the first
signal to arrive at the chamber amplifier/discriminator
was included with simulated data. (Our chamber TDC's only
registered the first hit received by each channel.)

Chamber inefficiencies were modeled in some
detail, since they
determined the characteristics of the
chamber component of the charged mode trigger.
Individual wire efficiencies were $\sim$99\%; the characteristics of
especially inefficient channels observed in the data were included in the
simulation.

The chamber simulation did not model
$\delta$-ray production.
Consequently, the high-side tail in the
track $\chi^2$ distribution is not simulated.
We expected that the effects of $\delta$-rays were not correlated
significantly
with the decay vertex $z$ position, and would only cause
an overall loss of data relative to Monte Carlo
without introducing biases in our phase-of-$\eta$
measurements.

\subsection{Lead glass simulation}
\label{sec:glass_MC}
The mean
amount of \v Cerenkov light which reached an individual block's
photomultiplier tube was a function of the incident particle's
energy,
the block's attenuation length for \v Cerenkov light, and the
depth at which the shower began.
Since showers were almost fully
contained in the calorimeter, signals from
showers which began deep in a block suffered less attenuation than
those which began early.
We parametrized the distribution about the mean
for the light reaching the phototube
as a Gaussian with high-side and low-side exponential tails.
Photon
showers were simulated as pairs of
electron showers which begin where the photon converted
to an $e^+e^-$ pair.
The use of an effective block length allowed us to
take into account
the fact photons began to shower
at different depths in the calorimeter.
Extensive studies using
EGS4\cite{EGS4} have shown this to be a sensible description
of the production of \v Cerenkov light in lead glass.\cite{JRP}

Studies with the xenon calibration flasher were used to
determine the mean number of photoelectrons per \v Cerenkov photon
expected for each
block.
The simulation chose the amount of \v Cerenkov
light which reached each phototube,
determined the expected number of photoelectrons,
then varied this assuming a Gaussian distribution whose width
was the square root of the expected mean.

The event simulation modeled cluster shapes using a
library of electron clusters
obtained from special runs.
In these calibration runs, a beam
of photons was converted to $e^+e^-$ pairs; the $e^+$ and $e^-$
were steered by calibration magnets (see Fig.~\ref{e773detector})
into the lead glass. (By varying the currents in the magnets we
were able to
illuminate the entire calorimeter with a wide range of electron
energies.)
Clusters were selected from the
library based on incident electron energies and positions
in the central block.
The signal in each block was scaled by a constant to give
the simulated response. For photons, two appropriate electron
clusters were overlayed.

We adjusted the tail structure in the glass response
to correct for effects
not modeled in the EGS4
simulation. For example, EGS4 assumed that the attenuation
of \v Cerenkov light was
uniform along the length of a block.
In reality, absorption increased with depth: there
was more radiation damage deep in blocks due to hadronic
interactions. In addition, the curing lamp (which was placed near the
upstream face of the calorimeter)
was less effective
at correcting damage in the downstream end of blocks.

\subsection{Veto and hodoscope simulation}
\label{sec:veto_MC}
Efficiencies of scintillation counters were measured using
muon tracks. Small gaps between adjacent counters in hodoscope planes
were mapped using the drift chambers and included in the Monte Carlo.
Since we had determined the
response of
photon veto counters with data we were able to
simulate their response to photons in the Monte Carlo.

\subsection{Comparisons of data and Monte Carlo}
\label{sec:dvsmc}
The large samples of $K \rightarrow \pi e \nu$ and
$K \rightarrow 3\pi^0$ decays provided us with useful tools
for verifying the accuracy of the Monte Carlo simulation.
We collected roughly five times as many $3\pi^0$ decays as
$2\pi^0$ decays, and more than ten times as many $K_{e3}$'s as
$\pi^+\pi^-$ decays.
It was particularly important that we modeled correctly
acceptance and resolution effects that influenced the
reconstructed vertex $z$ distributions.  Shown in
Figs.~\ref{pi0pi0pi0z} and \ref{ke3z} are
comparisons of these distributions
for data and Monte Carlo.
Our success in simulating the acceptance of these
``high statistics" modes gives us confidence in
the validity of conclusions we draw from characteristics of
the $\pi \pi$ and $\pi \pi \gamma$ samples.

\section{Analysis and Fits}
\label{sec:analysis}
\subsection{$\phi_{+-}$, $\Delta m$, $\tau_S$ analysis}
\label{sec:RAB}
We used the Monte Carlo detector simulation to predict the  $\pi^+\pi^-$
acceptance
in 10~GeV/c $\times$ 2~m cells in momentum $p$ and decay position $z$.
The acceptance in each cell was defined as the
ratio of the number of detected events with {\underline
{reconstructed}} $p$, $z$ in the cell divided by the number of generated
events with {\underline {true}} $p$, $z$ in that cell.
Since the simulation accurately modeled the data, effects due to resolution
smearing were properly included in our fits.
Average acceptance was approximately 20\%.

We measured $Arg(\eta_{+-})$ as well as the regeneration parameters
$\alpha$ and
$\left[\left(f - {\overline f}\right)/k\right]_{70~GeV}$
with a fit to background-subtracted data
which included
free parameters describing the beam intensities and energy
spectra. In this fit
we constrained the $K_L - K_S$ mass difference
to the value measured by E731:
$\Delta m = 0.5286 \times 10^{10} \hbar s^{-1}c^{-2}$ (Ref.~\cite{gibbons1}),
while using the world-average
value for the
$K_S$ lifetime:
$\tau_S = 0.8922 \times 10^{-10}s$ (Ref.~\cite{PDG}).
(These values correspond to $\phi_\epsilon = 43.33^\circ$.)
We found that $\phi_{+-} = (43.35 \pm 0.70)^\circ$ where the error
is statistical (systematic uncertainties will be discussed below).
The fit's $\chi^2$ was 598 for 577 degrees of freedom.

In a separate fit we determined $\Delta m$ and $\tau_S$
(as well as the beam and regeneration parameters)
by constraining the phase of $\eta_{+-}$ to equal the phase of $\epsilon$:
$$\phi_{+-} \equiv \phi_\epsilon =
tan^{-1}\left(2 \Delta mc^2 \tau_S / \hbar \right)\ .$$
We found that
$\Delta m = (0.5286 \pm 0.0029)\times 10^{10}\hbar s^{-1}c^{-2}$
and
$\tau_S = (0.8929 \pm 0.0014) \times 10^{-10}s$, corresponding to
$\phi_\epsilon = 43.35^\circ$. Again, the errors are statistical.
When $\phi_{+-}$ was unconstrained, we found that
$\phi_{+-} - \phi_\epsilon =  (-0.84 \pm 1.42)^\circ$,
$\Delta m = (0.5268 \pm 0.0041)\times 10^{10}\hbar s^{-1}c^{-2}$,
and
$\tau_S = (0.8942 \pm 0.0026) \times 10^{-10}s$.

Sources of systematic errors included uncertainties in
our determination of
experimental acceptance,
errors associated with background subtractions,
uncertainties in our descriptions of the two regenerators, and
possible deviations of the momentum dependence of
$\left(f - {\overline f}\right)/k$
from a power law.
We studied acceptance issues by moving the edges of limiting apertures
and introducing spurious $z-$dependent slopes (some of which were functions
of momentum) to force disagreement between data and Monte Carlo.
Background systematics were investigated by varying the
$p_T^2$ cut, thus changing the amount of
$K_{e3}$ contamination remaining with the coherent signal.
We changed the positions and relative lengths of the regenerators in the
apparatus simulation to determine our sensitivity to
uncertainties in regenerator geometry.
Issues concerning an unexpected momentum dependence in the
regeneration amplitude are described in Ref.~\cite{gibbonsthesis}.
The sensitivity of $\phi_{+-}$ to these effects is summarized in
Table~\ref{pipisystem}; combining the errors in quadrature yielded a net
systematic uncertainty of $\pm 0.79^\circ$.
The fit for $\phi_{+-}$ did not allow $\Delta m$ or $\tau_S$ to vary.
Increasing the values of $\Delta m$ and $\tau_S$ by one
(Particle Data Group) standard
deviation (Ref.~\cite{PDG}) would change $\phi_{+-}$ by $+0.38^\circ$ and
$-0.62^\circ$ respectively.

In summary, our preliminary results are
\begin{eqnarray}
&&\phi_{+-}  =
\left[ 43.35\ \pm\ 0.70\ (stat.)\ \pm\ 0.79\ (syst.)\right] ^\circ
\nonumber \\
&&(\Delta m {\rm\ fixed\ at\ }0.5286 \times 10^{10} \hbar s^{-1}c^{-2},
\ \tau_S  {\rm\ fixed\ at\ }0.8922 \times 10^{-10}s.) \nonumber \\
&&\delta(\phi_{+-}) = +0.38^\circ {\rm\ for\ } \delta(\Delta m) = 0.0024
\nonumber \\
&&\delta(\phi_{+-}) = -0.62^\circ {\rm\ for\ } \delta(\tau_S) = 0.0020\ .
\nonumber
\end{eqnarray}
If instead we constrain
$\phi_{+-} \equiv tan^{-1}(2\Delta mc^2 \tau_S / \hbar)$, we find
\begin{eqnarray}
&&\Delta m = \left[ 0.5286\ \pm\ 0.0029\ (stat.)\ \pm\ 0.0022\ (syst.)\right]
\times 10^{10}\hbar s^{-1}c^{-2}
\nonumber\\
&&\tau_S = \left[ 0.8929\ \pm\ 0.0014\ (stat.)\ \pm\ 0.0014\ (syst.)\right]
\times 10^{-10}s \ .
\nonumber
\end{eqnarray}
Fitting simultaneously for $\phi_{+-}$, $\Delta m$, and $\tau_S$ yields
\begin{eqnarray}
&&\phi_{+-} - \phi_\epsilon =  \left[-0.84 \pm 1.42\ (stat.)\  \pm
 1.22\ (syst.)\right]^\circ \nonumber \\
&&\Delta m = \left[0.5268 \pm 0.0041\ (stat.)\  \pm
 0.0029\ (syst.)\right]\times 10^{10} \hbar s^{-1}c^{-2}
\nonumber \\
&&\tau_S = \left[0.8942 \pm 0.0026\ (stat.)\  \pm
 0.0018\ (syst.)\right] \times 10^{-10}s\ .\nonumber
\end{eqnarray}
Systematic uncertainties will
decrease with additional analysis work.

These results can be compared with the Particle Data Group's world average
values\cite{PDG} and
previous measurements from
E731\cite{gibbons1} and NA31\cite{carosi90}.
The PDG values are
$\phi_{+-} =  \left( 46.5 \pm 1.2 \right)^\circ$,
$\Delta m = \left( 0.5351 \pm 0.0024 \right)\times 10^{10} \hbar s^{-1}c^{-2}$,
$\tau_S = \left(0.8922 \pm 0.0020 \right) \times 10^{-10}s$.
Since the value of $\phi_{+-}$ is correlated with the values of $\Delta m$
and $\tau_S$, the PDG used the world-average values of $\Delta m$
and $\tau_S$ when combining $\phi_{+-}$ results from different experiments.
If we were to recalculate our $\phi_{+-}$ result using
the PDG values for $\Delta m$
and $\tau_S$ we would obtain $\phi_{+-} = (44.38 \pm 0.70 \pm 0.79)^{\circ}$.
The E731 results were
$\phi_{+-} =  \left( 42.2 \pm 1.5 \right)^\circ$,
$\Delta m = \left( 0.5286 \pm 0.0028 \right)\times 10^{10} \hbar s^{-1}c^{-2}$,
$\tau_S = \left(0.8929 \pm 0.0016 \right) \times 10^{-10}s$.
The result for $\phi_{+-}$ was obtained from fits which
assumed $\tau_S = 0.8922 \times 10^{-10}$ but allowed $\Delta m$ to float.
The results for $\Delta m$ and $\tau_S$ fixed $\phi_{+-}=\phi_{\epsilon}$.
If we were to
recalculate our $\phi_{+-}$ result using the $\Delta m$
value obtained\cite{gibbonsthesis} in E731's $\phi_{+-}$ fit, we would find
$\phi_{+-} = (42.14 \pm 0.70 \pm 0.79)^{\circ}$.
The NA31 result was determined assuming
the PDG values $\tau_S = 0.8922 \times 10^{-10}$
and $\Delta m = 0.5351 \times 10^{10}$. They found
$\phi_{+-} =  \left( 46.9 \pm 2.2 \right)^\circ$.
This can be compared with our $\phi_{+-}$ result recalculated
using the PDG values for
$\Delta m$ and $\tau_S$
mentioned above.
Our results are consistent with the absence of CPT violation in
$K_L \rightarrow \pi^+ \pi^-$ decays.

\subsection{$\Delta \phi$ analysis}
\label{sec:BS}
Our preliminary result for $\phi_{00} - \phi_{+-}$ is based on
data
written after the T,V hodoscope was removed,
approximately 70\% of the full data set.
As was the case in the
$\phi_{+-}$ analysis, we
binned data and Monte Carlo samples for both $\pi^0\pi^0$ and $\pi^+\pi^-$
decays
in 10~GeV/c $\times$ 2~m $(p,z)$ cells.  The copious
$3\pi^0$ sample served to verify the accuracy of the neutral mode
Monte Carlo.
Acceptance for neutral mode decays varied with
position and $K$ energy, and was typically $\sim 5$\%.

We fit our background-subtracted
data for values of $\Delta \phi$, the regeneration parameters
$\alpha$ and
$\left[\left(f - {\overline f}\right)/k\right]_{70~GeV}$, $\phi_{+-}$,
$\epsilon'$, and beam intensity and energy spectrum parameters.
For the cuts described in Table~\ref{deltaphicuts}, the fit
$\chi^2$ was 682 for 619 degrees of freedom.
We found that $\Delta \phi$ returned by the
fit varied somewhat with the value of the minimum cluster energy cut,
$E_{cl\ min}$.
For our preliminary result we report the average
of two fits which used $E_{cl\ min}=2.2~GeV$ and
$E_{cl\ min}=4.0~GeV$, and assign a systematic uncertainty of $0.5^\circ$
to this effect.
We find $\Delta \phi = (0.67 \pm 0.85)^\circ$, where the error
is statistical.

Other systematic uncertainties arose from
possible inaccuracies in our determination of
acceptance,
our background subtractions,
and our parametrization of the lead glass calorimeter.
Effects involving regeneration influenced
both $\phi_{00}$ and $\phi_{+-}$, and canceled when
extracting their difference.
As was the case with the charged mode analysis, we
introduced spurious $z-$dependent slopes
to force disagreement between data and Monte Carlo in order
to gauge the consequences of acceptance errors.
We varied the amounts of background from
$3\pi^0$ decays, inelastically and diffractively regenerated
$K_S$, and from beam interactions in material in the detector.
We studied several possible sources of calorimeter misparametrization,
including an energy scale mismatch between the lead glass and the
tracking system, misrepresentation of the lead glass' energy
resolution, and uncorrected pedestal shifts.
The dependence of $\Delta \phi$ on these effects is summarized in
Table~\ref{pi0pi0system}; the quadrature sum of the various
sources of systematic
uncertainty is $\pm 1.1^\circ$.
Consequently, our preliminary result for the phase difference between
$\eta_{00}$ and $\eta_{+-}$ is
\begin{eqnarray}
&&\Delta \phi = \left[ 0.67\ \pm\ 0.85\ (stat.)\ \pm\ 1.1\ (syst.)\right]
^\circ \ .
\nonumber
\end{eqnarray}
We expect the statistical error to
decrease when the full data set is used
in the analysis; the systematic error will
decrease as the analysis continues.

We can compare our result with the Particle Data Group's world average
value\cite{PDG} and
previous measurements from
E731\cite{gibbons1} and NA31\cite{carosi90}.
The 1992 PDG value is
$\Delta \phi =  \left( -0.1 \pm 2.0 \right)^\circ$.
The E731 result of
$\Delta \phi =  \left( -1.6 \pm 1.2 \right)^\circ$
was determined with a fit which allowed the values for
$\Delta m$ and $\tau_S$ to vary.  If E731 had calculated their
result using our $\Delta m$ and $\tau_S$ values they
would have obtained
$\Delta \phi =  -1.75^\circ$.
The NA31 result was
$\Delta \phi =  \left( 0.2 \pm 2.9 \right)^\circ$, and was
determined assuming
the PDG values $\tau_S = 0.8922 \times 10^{-10}$
and $\Delta m = 0.5351 \times 10^{10}$.
Their value is insensitive to changes in $\Delta m$.
Our result is consistent with previous measurements and
with the absence of CPT violation in
$K_L \rightarrow \pi \pi$ decays.

\subsection{$\eta_{+-\gamma}$ analysis}
\label{sec:JM}
The Monte Carlo simulation of the apparatus was used
to determine the acceptance as a function of $p$ and $z$ for
$\pi \pi \gamma$ decays in each beam.
We generated simulated data sets which were approximately twenty times
larger than the corresponding real data sets and
determined average acceptance in
each cell of the $(p,z)$ grids described above.
Overall acceptance  of our apparatus for the $\pi^{+}\pi^{-}\gamma$
decay mode was typically 11\%.
The momentum spectrum and flux of kaons striking the regenerators
were determined from the more copious
$K \rightarrow \pi^{+}\pi^{-}$ decay mode.
Characteristics of the regeneration amplitude
$\rho$ were also determined with $\pi^{+}\pi^{-}$ data.
The $\pi^+ \pi^- \gamma$ analysis (and Monte Carlo)
assumed that
$\Delta m = 0.5286 \times 10^{10} \hbar s^{-1}c^{-2}$ (Ref.~\cite{gibbons1})
and that the  $K_{S} \rightarrow \pi^{+}\pi^{-}\gamma$
branching ratio (Ref.~\cite{ramberg1})
was $4.87 \times 10^{-3}$.

We extracted phase information about $\eta_{+-\gamma}$
from the interference between
the inner bremsstrahlung
$K_{L,S} \rightarrow \pi \pi \gamma$ decay amplitudes.
The direct emission amplitude (which tended to produce a more
energetic photon in the kaon rest frame) only participated in the $K_L$ decay
rate, and did not interfere with the $K_S$ decay amplitude
at an observable level.
Based on a previous measurement (Ref.~\cite{ramberg1}),
the Monte Carlo simulation assumed that
the direct emission contribution
to the $K_{L}$ decay rate with $E_{\gamma}^{*} >$ 20~MeV
was 68.5\%. Here,  $E_{\gamma}^{*}$ is the photon energy in the
$K$ rest frame.

The four background-subtracted $(p,z)$ data grids were
fit simultaneously
for the magnitude and phase of $\eta_{+-\gamma}$, the $K_S \rightarrow
\pi^+\pi^-\gamma$ branching ratio, and the ratio of the
direct emission and inner bremsstrahlung contributions to
the $K_L$ decay amplitude
using a log-likelihood procedure.
Our
preliminary result is
$|\eta_{+-\gamma}| = (2.414 \pm 0.065) \times 10^{-3}$
and
$\phi_{+-\gamma} \equiv Arg(\eta_{+-\gamma})= (45.47 \pm 3.61)^{\circ}$,
where the errors are statistical.
After fitting, a $\chi^{2}$ was calculated using all bins with at
least five events. We found $\chi^{2} = 366.9$ for 347 degrees of freedom.
Results from the individual data sets and the combined sample
are summarized in Table~\ref{jmsets}.

Possible sources of systematic error included uncertainties associated with
the regeneration amplitude, background subtractions, and our
parametrization of the kaon flux and momentum
spectrum.  In addition, our
descriptions of $K_L \rightarrow \pi \pi \gamma$ direct emission and the
$K_S \rightarrow \pi \pi \gamma$ branching ratio
contributed to the
systematic error.  By varying the analysis parameters by one standard
deviation,
we studied
the extent to which systematic uncertainties influenced our conclusions.
The results are presented
in Table \ref{system}.
Adding systematic errors in quadrature yields the preliminary results
\begin{eqnarray}
&&|\eta_{+-\gamma}| = \left[2.414\ \pm\ 0.065\ (stat.)\ \pm
\ 0.062\ (syst.)\right] \times 10^{-3} \nonumber \\
&&\phi_{+-\gamma}  =  \left[ 45.47\ \pm\ 3.61\ (stat.)\ \pm
\ 2.40\ (syst.)\right] ^\circ \ .
\nonumber
\end{eqnarray}
These errors are significantly smaller than those of
the best previous measurement (Ref.~\cite{ramberg2}) which found
$|\eta_{+-\gamma}| = (2.15 \pm 0.26 \pm 0.17)\times 10^{-3}$ and
$\phi_{+-\gamma} = (72 \pm 23 \pm 17)^{\circ}$.
Our value for $\eta_{+-\gamma}$, and the best previous measurement,
are compared in Fig.\ref{etappgplot}.
This measurement of $\eta_{+-\gamma}$ is also not significantly different
from the analogous $\pi^{+}\pi^{-}$ amplitude ratio
$|\eta_{+-}| = (2.268 \pm 0.023)\times 10^{-3}$ (Ref.~\cite{PDG})
and the
$\phi_{+-}$ value presented in Sec.~\ref{sec:RAB}.
This result is consistent with the absence of CP violation beyond that
present in $K_L \rightarrow \pi \pi$ decays.

\section{Conclusions}
\label{sec:conclusions}
We report preliminary results from analysis of data from the
1991 fixed target run at Fermilab. Our results for $Arg(\eta_{+-})$,
$\Delta m$, $\tau_S$, $\vert \eta_{+-\gamma} \vert$, and $Arg(\eta_{+-\gamma})$
are based on the entire data set. Our result for
$\Delta \phi \equiv Arg(\eta_{00})-Arg(\eta_{+-})$ is based on analysis
of about 70\% of the available data. Systematic uncertainties will
decrease as we refine the analyses; the statistical precision
of our $\Delta \phi$ result will improve when the full data set
is used.

Our two-pion
results are consistent with CPT conservation in $K \rightarrow \pi \pi$
decays. Our $\eta_{+-\gamma}$ result is consistent with the absence
of an unusual source of CP violation in $K \rightarrow \pi \pi \gamma$
decays.

%
\appendix
\section*{Regeneration}
\label{sec:regeneration}
\subsection{$K_S$ coherent regeneration}
\label{sec:kscohregeneration}
Neglecting decays, the wave function
for a relativistic $K^0$ meson with wavenumber $k$
will obey the Klein-Gordon equation
$(\nabla^2 + k^2)\psi_{K}({\bf x}) = 0$.
If the $K^0$ passes through a scattering
medium,
its total wavefunction will become the sum of the incident and scattered
waves.
Defining f(k) to be the average forward scattering amplitude at
each scattering site inside the medium, we find that\cite{foldy}
$\psi_K$ must satisfy
\[
(\nabla^2 + k^2 + 4\pi Nf(k))\,\psi_{K} = 0
\]
with approximate solution
\begin{equation}
\psi_K({\bf x},t) \approx
e^{i\left[(1+{{2\pi N f} \over {k^2}})kz\ -\
Et/\hbar\right]}
\label{psievolution}
\end{equation}
for a plane wave traveling in the $z$ direction.
Here, $N$ is the number of scatterers per unit volume.
Similarly, the
wavefunction for a ${\overline {K^0}}$ with forward scattering amplitude
${\overline {f}}(k)$ will be
\begin{equation}
\psi_{\overline {K}}({\bf x},t) \approx
e^{i\left[(1+{{2\pi N {\overline {f}}(k)} \over {k^2}})kz
\ -\ Et/\hbar\right]}\ .
\label{psibarevolution}
\end{equation}
Note that these expressions include the effects of elastic
and inelastic scattering since
a positive imaginary part to $f(k)$ (or ${\overline {f}}(k)$)
causes $\vert\psi\vert$ to decrease.
This is sensible:  the optical theorem guarantees
$Im f > 0$ whenever the total scattering cross section is
nonzero.

We can use Eqs.\ (\ref{psievolution}, \ref{psibarevolution})
to describe the position dependence
of a mixed state traveling in the $z$ direction through a regenerator.
The $z$-dependence of the $K_S$, $K_L$ amplitudes for the state
$\vert K \rangle\ =\ A_S\vert K_S \rangle\ + A_L\vert K_L \rangle$
will obey
\begin{eqnarray}
{{\partial A_S} \over {\partial z}}\ &=&\ \left[ik + iN\pi
\left({{f + {\overline f}}\over {k}}\right)\right]A_S
\ \ +\ \ iN\pi \left({{f - {\overline f}}\over {k}}\right)A_L
\phantom{\ ,} \nonumber
\\
{{\partial A_L} \over {\partial z}}\ &=&\ \left[ik + iN\pi
\left({{f + {\overline f}}\over {k}}\right)\right]A_L
\ \ +\ \ iN\pi \left({{f - {\overline f}}\over {k}}\right)A_S
\ ,\nonumber
\end{eqnarray}
neglecting weak interactions and decays.
A state which is pure $K_L$ at $z=0$ will evolve into a mixture
of $K_L$ and $K_S$
as long as $\left(f - {\overline f}\right) \neq 0$.
In particular, the ratio of the $K_S$ and $K_L$ amplitudes a small distance
$\Delta z$ inside a regenerator will be
\begin{equation}
\rho \equiv {{A_S(\Delta z)} \over {A_L(\Delta z)}} \approx iN\pi
\left({{f - {\overline f}}\over {k}}\right)\Delta z \ . \label{rhocoher}
\end{equation}
This {\it coherent regeneration amplitude} $\rho$
results from constructive interference among all the outgoing spherical waves
produced at each of the scattering sites in the regenerator.
Since the wavelength of a 100~GeV/c $K$ meson is $\sim 2 \times
10^{-18}$ meters, coherent scattering in a macroscopic
regenerator only takes place for extremely small scattering angles.

The total cross section at high energy is dominated by inelastic
processes so the scattering amplitudes $f$ and ${\overline f}$ are nearly
imaginary\cite{goldberger}.
The momentum-dependence of $\vert (f - {\overline f})/k \vert$
has been measured\cite{gibbonsthesis} and is in
agreement with the
Regge theory prediction\cite{gilman} that
\begin{equation}
\left({{f - {\overline f}} \over k}\right) \ \propto
\ p^{\alpha}e^{-i \pi (2+\alpha)/2}\ .
\end{equation}
The connection between $\alpha$ and the phase of $f-{\overline f}$ comes
from analyticity, and does not depend on
the validity of Regge theory.
\subsection{Diffraction regeneration}
\label{sec:diffraction}
Scattering does not mix
$K^0 \leftrightarrow {\overline {K^0}}$.
However, since the
scattering probability for a $K^0$
differs from that for a ${\overline {K^0}}$,
a $K$ meson which contains a mix of $K^0$ and ${\overline {K^0}}$
amplitudes
will scatter into a state containing a different combination
of $K^0$ and ${\overline {K^0}}$.
In terms of the elastic scattering amplitudes $f_k(\theta,\phi)$ and
${\overline f}_k(\theta,\phi)$,
the state
\[
\psi_{i} \equiv
A\vert K^0 \rangle\ + {\overline A}\vert {\overline {K^0}} \rangle
\]
will scatter elastically into the state
\[
\psi_{f}\ =
\ \left[A\,f_k(\theta,\phi)\vert K^0 \rangle\ + {\overline A}\,
{\overline f}_k(\theta,\phi)
\vert {\overline {K^0}} \rangle\right]{{e^{ikr}} \over r}\ .
\]
For convenience, we define
$f_{22}(\theta,\phi) \equiv f_k(\theta,\phi) + {\overline f}_k(\theta,\phi)$
and
$f_{21}(\theta,\phi) \equiv f_k(\theta,\phi) - {\overline f}_k(\theta,\phi)$.

If we rewrite the incident and final states $\psi_i$,
$\psi_f$ in terms of $K_S$ and $K_L$,
we have
\begin{eqnarray}
\psi_{i}&\ =\ &A_S\vert K_S \rangle\ + A_L\vert K_L \rangle  \label{psii} \\
\psi_{f}&\ =\ &
\Biggl\{
\Bigl( A_S f_{22} + A_L f_{21} \Bigr) \vert K_S \rangle
\ \ + \ \
\Bigl( A_L f_{22} + A_S f_{21} \Bigr) \vert K_L \rangle
\Biggr\}
{{e^{ikr}} \over {2r}}\ . \label{psif}
\end{eqnarray}
In the final state $K_S$ amplitude, the term
$A_S f_{22}$
corresponds to elastic $K_S \rightarrow K_S$ scattering while the term
$A_L f_{21}$
comes from (regenerative) $K_L \rightarrow K_S$ scattering.
Similarly, in the final $K_L$ amplitude,
the term
$A_L f_{22}$
arises from elastic $K_L \rightarrow K_L$ scattering, while
$A_S f_{21}$
results from $K_S \rightarrow K_L$ scattering.

A pure $K_L$
will scatter into a mixed
$K_L$, $K_S$ state; we can define the {\it diffraction regeneration
amplitude} $\rho_D$ for elastic scattering into $\theta,\phi$ as the
ratio
$f_{21} / f_{22}$.
With this definition
we can rewrite Eq.~(\ref{psif}) as
\begin{eqnarray}
\psi_{f}&\ =\ &
\Biggl\{
\Bigl( A_S  + \rho_D A_L \Bigr) \vert K_S \rangle
\ \ + \ \
\Bigl( A_L + \rho_D A_S \Bigr) \vert K_L \rangle
\Biggr\}
f_{22}{{e^{ikr}} \over {2r}}\ . \label{psif2}
\end{eqnarray}
Because
$f_k$ and ${\overline f}_k$ are nearly
imaginary (Ref.\ \cite{goldberger}),
$f_{22} \approx i\vert f_{22} \vert$
so that
\[
\rho_D(\theta,\phi)\ =\ {{f_{21}} \over {f_{22}}}
\ \approx
\ -i{{f_{21}} \over
{\vert f_{22} \vert}} \ .
\]
With the help of the optical theorem, this can be rewritten as
\begin{equation}
\rho_D(\theta,\phi)\ \approx\ {{-i2\pi} \over {\sigma_T}}
\left({{f_{21}} \over k}\right) \label{rhodiff2}
\end{equation}
when $\theta,\phi$ are close to
the forward direction so that
$f_{22}(\theta, \phi) \approx f_{22}(0,0)$.
Here, $\sigma_T$ is the $K_L$-nucleus
total scattering cross section.

There are interference effects associated with the presence of both
coherent and diffractive processes.
Imagine that a $K_L$ enters a regenerator of
length $L$, scatters elastically after traveling a distance
$\Delta z$, then leaves the regenerator without rescattering.
(We will ignore effects associated
with finite regenerator length such as the
decay of the $K_S$ amplitude and the
phase drift between
the kaon's $K_S$ and $K_L$ components due to the
non-zero $K_L-K_S$ mass difference.)
{}From Eq.~(\ref{rhocoher}), the kaon's wavefunction immediately
before scattering will be
\begin{eqnarray}
\psi_{before}&\ =\ &
 \left({{iN\pi \Delta z f_{21}(0,0)}\over {k}}\right)
\vert K_S \rangle
\ \ +\ \ \vert K_L \rangle\ ,  \label{psii2}
\end{eqnarray}
neglecting normalization. The $K_S$ component is the result of coherent
regeneration.
Immediately after scattering, the kaon's wavefunction will become
\begin{eqnarray}
\psi_{after}&\ =\ &
\left\{
\left( {{iN\pi \Delta z f_{21}(0,0)}\over {k}} + \rho_D \right)
\vert K_S \rangle
\ \ + \ \
\left( 1 + {{iN\pi \Delta z f_{21}(0,0)}\over {k}} \rho_D \right)
\vert K_L \rangle
\right\}
f_{22}{{e^{ikr}} \over {2r}}\ . \label{psif3}
\end{eqnarray}
The first term in the $K_S$ amplitude corresponds to $K_S \rightarrow K_S$
scattering while $\rho_D$ corresponds to
elastic $K_L \rightarrow K_S$ scattering.
For the E773 regenerator geometry and $K$ energy range, both
contributions to
the $K_S$ amplitude of Eq.~(\ref{psif3}) are small.
The modification to the $K_L$ amplitude is second-order in
the regeneration parameters $f_{21}$, $\rho_D$.
As the kaon travels through the remaining distance $(L-\Delta z)$
in the regenerator,
its $K_S$ and $K_L$ amplitudes will continue to evolve because of
$K_{S,L} \rightarrow K_{L,S}$ and $K_{L,S} \rightarrow K_{L,S}$
forward scattering.
In particular, its $K_S$ amplitude will
gain an additional contribution from coherent regeneration.
Neglecting normalization, and terms
which are second order in the regeneration parameters,
we find that the kaon's wavefunction at the downstream end of the regenerator
is
\begin{eqnarray}
\psi_{f}&\ =\ &
\left\{
\left( {{iN\pi L f_{21}(0,0)}\over {k}} + \rho_D \right)
\vert K_S \rangle
\ \ + \ \
\vert K_L \rangle
\right\}
f_{22}{{e^{ikr}} \over {2r}}\ . \label{psif4}
\end{eqnarray}
With the definition of the coherent regeneration amplitude $\rho$
in Eq.~(\ref{rhocoher}), we can rewrite Eq.~(\ref{psif4}) as
\begin{eqnarray}
\psi_{f}&\ =\ &
\Bigl[
\left( \rho + \rho_D \right) \vert K_S \rangle
\ \ + \ \
\vert K_L \rangle
\Bigr]
f_{22}{{e^{ikr}} \over {2r}}\ . \label{psif5}
\end{eqnarray}

Because of
the relative minus sign between
$\rho$ and $\rho_D(\theta,\phi)$, there will be destructive interference
between coherent and diffractive regeneration.
If we approximate $f_{21}(\theta, \phi) \approx f_{21}(0,0)$
and make use of Eqs.~(\ref{rhocoher}) and (\ref{rhodiff2}),
we find
\begin{equation}
\rho + \rho_D \ \approx \ \rho \left( 1 - {2 \over {N L \sigma_T}} \right)\ .
\label{rhorhod}
\end{equation}
For a two interaction length regenerator, $NL\sigma_T = 2$,
and the destructive interference will be nearly perfect.

The interference between the coherent and diffractive regeneration
amplitudes is influenced by a number of factors in addition to regenerator
length.
If
$f_k(\theta,\phi)$ and ${\overline f}_k(\theta,\phi)$
are not purely imaginary, Eq.~(\ref{rhodiff2}) will be
inaccurate and the cancellation
will be imperfect.
To the extent that dependence on momentum transfer spoils the approximation
$f_{22}(\theta, \phi) \approx f_{22}(0,0)$,
Eq.~(\ref{rhorhod}) will be unreliable.
The nonzero $K_L$-$K_S$ mass difference
makes the $K_S$
amplitude's phase lag behind the $K_L$ phase as the kaon travels through the
regenerator: the $K_S$
amplitude from elastic scatters at the upstream and downstream ends of the
regenerator will contribute with slightly different phases to the decay
amplitudes. This phase drift is negligible at Fermilab energies, where
it is $\sim 5^\circ$ per meter of flight path for a 100 GeV kaon.
In addition, the finite $K_S$ lifetime
plays a role. The contribution to the $K_S$ amplitude from coherent
forward scattering will decay in flight as the kaon moves
through the regenerator.  The fraction which survives is a function of
the regenerator length and the kaon's lifetime, and is
independent of the location of the elastic scatter.
The diffractive contribution to the $K_S$ amplitude,
created at the scattering site,  will
also decay in flight. As a result, the
surviving diffractive contribution depends on the distance from
the scattering site to the downstream end of the regenerator.
Consequently, the sum of the coherent and diffractive $K_S$ amplitudes
will vary slightly with the location of the scattering site.

\bigskip
\bigskip
\begin{center}
\bf
ACKNOWLEDGEMENTS \\
\end{center}
\bigskip
This work was supported in part by Department of Energy grants
DE-FG0291ER-40677 and DE-FGO290ER-40560,
Department of Energy contract DE-ACO2-76CH03000, and by
National Science Foundation grants NSF-PHY-9406402,
NSF-PHY-9115027, and NSF-PHY-9019706.
%

%
%
%
%
\begin{table}
\addcontentsline{toc}{section}{TABLES}
\caption{Requirements on events
used in $\pi^+\pi^-$ and $\pi^{+}\pi^{-}\gamma$ analyses.
Set 1 (2) data were recorded before (after) removal of the T,V hodoscope.}
\label{cuts}
\begin{tabular}{lll}
                       &$\pi^+\pi^-$  &$\pi^{+}\pi^{-}\gamma$ \\
\tableline
Invariant mass, ${\rm MeV/c^2}$  &  $484 \leq m_{\pi \pi} \leq 512$
     &  $484 \leq m_{\pi \pi \gamma} \leq 512$ \\
${\rm p_T^2,\ (MeV/c)^2}$ &  $\leq 250$   &  $\leq 150$   \\
track momentum, GeV/c     & ${\rm \geq 7}$   &   ${\rm \geq 7}$  \\
$E/p$     & ${\rm \leq 0.8}$   &   ${\rm \leq 0.85}$  \\
Kaon energy (GeV) & $30 \leq E_K \leq 160$ & $25 \leq E_K \leq 155$ \\
Decay vertex, upstr. reg. (m)
&$118.5 \leq z \leq 127$ (set 1)
&$117.33 \leq z \leq 139.33$ (set 1)\\
&.OR. $129 \leq z \leq 140$ & \\
&$118.5 \leq z \leq 127$ (set 2)
&$117.33 \leq z \leq 159.33$ (set 2)\\
&.OR. $129 \leq z \leq 154$ & \\
Decay vertex, dnstr. reg. (m)
&$130 \leq z \leq 140$ (set 1)
&$128.63 \leq z \leq 140.63$ (set 1)\\
&$130 \leq z \leq 154$ (set 2)
&$128.63 \leq z \leq 158.63$ (set 2)\\
\tableline
$\Lambda$ suppression
& $m_{p \pi} \leq 1110\ {\rm MeV/c^2}$
& $m_{p \pi} \leq 1100\ {\rm MeV/c^2}$ \\
& .OR. $m_{p \pi} \geq 1122\ {\rm MeV/c^2}$
& .OR. $m_{p \pi} \geq 1130\ {\rm MeV/c^2}$ \\
& 
& .OR. $E_{\Lambda} \leq 100$ GeV \\
& 
& .OR. $p_p / p_\pi < 3$ \\
\end{tabular}
\end{table}
%
\begin{table}
\caption{Estimate of backgrounds in $\pi^+\pi^-$ sample.
}
\label{pipibkgrd}
\begin{tabular}{lcc}
                       &set 1  & set 2 \\
\tableline
Upstream Regenerator    &    0.29\%              &   0.25\%       \\
\tableline
Downstream Regenerator &    0.90\%              &   0.75\%       \\
\end{tabular}
\end{table}
%
\begin{table}
\caption{Number of coherent $\pi^{+}\pi^{-}$ events.
} 
\label{pipievents}
\begin{tabular}{lccc}
                       &set 1  & set 2  & Whole Run \\
\tableline
Upstream Regenerator   &  707k   &    750k     &   1,457k   \\
Downstream Regenerator &  160k   &    207k     &   367k  \\
Both Regenerators      &  867k  &    957k     &   1,824k   \\
\end{tabular}
\end{table}
%
\begin{table}
\caption{Number of coherent $\pi^{+}\pi^{-}\gamma$ events
after background subtraction.}
\label{events}
\begin{tabular}{lccc}
                       &set 1  & set 2  & Whole Run \\
\tableline
Upstream Regenerator   &  4,559 $\pm$ 11
&    3,779 $\pm$ 9    &   8,338 $\pm$ 14  \\
Downstream Regenerator &  1,209 $\pm$ 6
&    1,222 $\pm$ 6    &   2,431 $\pm$ 9  \\
Both Regenerators      &  5,768 $\pm$ 13
&    5,001 $\pm$ 11    &  10,769 $\pm$ 17  \\
\end{tabular}
\end{table}
%
%
\begin{table}
\caption{Estimate of backgrounds in $\pi^+\pi^-\gamma$ sample.}
\label{bkgrd}
\begin{tabular}{lcc}
                       &set 1  & set 2 \\
\tableline
Upstream Regenerator   & 75.5 $\pm$ 11.0 events &  65.0 $\pm$ 9.1 events \\
                       &    1.6\%              &   1.7\%       \\
\tableline
Downstream Regenerator & 28.0 $\pm$ 6.0 events  &  31.9 $\pm$ 6.3 events \\
&    2.3\%         &      2.5\%       \\
\end{tabular}
\end{table}
%
\begin{table}
\caption{Requirements on events used in the $\Delta \phi$ analysis.
}
\label{deltaphicuts}
\begin{tabular}{lll}
                       &$\pi^0\pi^0$  &$\pi^{+}\pi^{-}$ \\
\tableline
Invariant mass (${\rm MeV/c^2}$)  &  $474 \leq m_{\pi \pi} \leq 522$
     &  $484 \leq m_{\pi \pi \gamma} \leq 512$ \\
${\rm p_T^2,\ (MeV/c)^2}$ &  ---   &  $\leq 250$   \\
ring number (${\rm cm^2}$) &  $\leq 120$   &  ---   \\
track momentum (GeV/c)     & ---   &   ${\rm \geq 7}$  \\
cluster energy (GeV)     &  $E_{cluster} \geq 2.2$   &   ---  \\
$E/p$     & ---   &   ${\rm \leq 0.8}$  \\
Kaon energy (GeV) & $40 \leq E_K \leq 150$ & $30 \leq E_K \leq 160$ \\
Decay vertex (m) & upstr. reg.: $120 \leq z \leq 152$
                 & upstr. reg.: $118 \leq z \leq 127$\\
                && .OR. $129 \leq z \leq 152$ \\
              & dnstr. reg.: $130 \leq z \leq 152$
               & dnstr. reg.: $129 \leq z \leq 152$ \\
\tableline
$\Lambda$ suppression
& ---
& $m_{p \pi} \leq 1110\ {\rm MeV/c^2}$ \\
& & .OR. $m_{p \pi} \geq 1122\ {\rm MeV/c^2}$ \\
\end{tabular}
\end{table}
%
\begin{table}
\caption{Number of coherent $\pi^{0}\pi^{0}$ events.
(Note that only
data set 2 is used in the preliminary result for $\Delta \phi$). }
\label{p0p0events}
\begin{tabular}{lccc}
                       &set 1  & set 2  & Whole Run \\
\tableline
Upstream Regenerator   &  85k   &    180k     &   265k   \\
Downstream Regenerator &  35k   &     75k     &   110k  \\
Both Regenerators      & 120k   &    255k     &   375k   \\
\end{tabular}
\end{table}
%
\begin{table}
\caption{Estimates of systematic error in $\pi^+\pi^-$ analysis.}
\label{pipisystem}
\begin{tabular}{lc}
Source of uncertainty & change in $\phi_{+-}$ \\
\tableline
Acceptance calculation
&$0.57^{\circ}$ \\
Analyticity constraint on phase of $\rho$
&$0.5^\circ$ \\
Background subtractions
&$0.2^\circ$ \\
Regenerator descriptions
&$0.05^\circ$ \\
Total     &$0.79^\circ$  \\
\tableline
Increase $\Delta m$ by 0.0024 (1 PDG $\sigma$)
& $\phi_{+-}$ increases $+0.38^\circ$ \\
Increase $\tau_S$ by 0.0020 (1 PDG $\sigma$)
& $\phi_{+-}$ decreases $-0.62^\circ$\\
\end{tabular}
\end{table}
%
\begin{table}
\caption{Estimates of systematic error in $\Delta \phi$ analysis.}
\label{pi0pi0system}
\begin{tabular}{lc}
Source of uncertainty & change in $\Delta \phi$ \\
\tableline
Acceptance parametrization for $\pi^+\pi^-$ and $\pi^0\pi^0$
&$0.3^{\circ}$ \\
Background subtractions
&$0.6^\circ$ \\
Calorimeter energy scale
&$0.6^\circ$ \\
Calorimeter nonlinearity (including minimum cluster energy cut)
&$0.5^\circ$ \\
Calorimeter resolution
&$0.5^\circ$ \\
\tableline
Total     &$1.1^\circ$  \\
\end{tabular}
\end{table}
%
\begin{table}
\caption{$\eta_{+-\gamma}$ magnitude and phase from different data sets.}
\label{jmsets}
\begin{tabular}{lll}
Regenerator &set 1  & set 2 \\
\tableline
Upstream
&$\vert \eta_{+-\gamma}\vert = (2.358 \pm 0.086) \times 10^{-3}$
&$\vert \eta_{+-\gamma}\vert = (2.406 \pm 0.084) \times 10^{-3}$
\\
&$Arg(\eta_{+-\gamma}) = (51.1 \pm 5.4)^\circ$
&$Arg(\eta_{+-\gamma}) = (40.9 \pm 6.5)^\circ$
\\
\tableline
Downstream
&$\vert \eta_{+-\gamma}\vert = (2.376 \pm 0.098) \times 10^{-3}$
&$\vert \eta_{+-\gamma}\vert = (2.443 \pm 0.083) \times 10^{-3}$
\\
&$Arg(\eta_{+-\gamma}) = (43.1 \pm 7.8)^\circ$
&$Arg(\eta_{+-\gamma}) = (54.0 \pm 6.5)^\circ$
\\
\tableline
\tableline
All data
&$\vert \eta_{+-\gamma}\vert = (2.414 \pm 0.065 \pm 0.062) \times 10^{-3}$
&$Arg(\eta_{+-\gamma}) = (45.5 \pm 3.6 \pm 2.4)^\circ$
\end{tabular}
\end{table}
%
\begin{table}
\caption{Estimates of systematic error in $\pi^+\pi^-\gamma$ analysis.}
\begin{tabular}{lccc}
Source of uncertainty & fractional change & change in $|\eta_{+-\gamma}|$
&change in in $\phi_{+-\gamma}$ \\
\tableline
$K_S \rightarrow \pi\pi\gamma$ ranching ratio
&2.3\% & $0.030 \times 10^{-3}$&$1.76^\circ$ \\
DE/IB ratio& 6.0\%  & $0.041\times 10^{-3}$ & $0.19^\circ$  \\
regeneration power law $\alpha$
& 2.5 \% & $0.001\times 10^{-3}$ &  $0.92^\circ$  \\
Normalization& 1.6\%   & $0.021\times 10^{-3}$ &  $1.22^\circ$  \\
Background normalization& 16\% & $0.008\times 10^{-3}$ &  $0.15^\circ$  \\
Background shape&    & $0.030\times 10^{-3}$ &  $0.17^\circ$  \\
Data/MC $p$ spectrum mismatch & $2\times10^{-4}$
& $0.002\times 10^{-3}$ & $0.54^\circ$  \\
Total     &          & $0.062\times 10^{-3}$ &  $2.40^\circ$  \\
\end{tabular}
\label{system}
\end{table}%
%
%
%
%
%
%
\begin{figure}
\addcontentsline{toc}{section}{FIGURES}
\caption{Elevation view of the E773 detector.  Kaons in the
beams travel to the right in the figure. Details of the individual
hardware systems appear in the text. The thin hodoscope labeled ``T,V"
was removed partway through the run.}
\label{e773detector}
\end{figure}
%
\begin{figure}
\caption{Neutral beam profiles
at the lead glass array, as determined with $K_L \rightarrow \pi^+ \pi^-$
decays.  Data are shown as the histogram with error bars; results
from the Monte Carlo simulation are shown as points.}
\label{kebeams}
\end{figure}
%
\begin{figure}
\caption{Predicted $K_L$ energy spectrum
striking the regenerators. The smooth curve
is based on the parametrization of $K^+$ and $K^-$
production by Malensek,
modified to agree with our data.}
\label{k_energy_spectrum}
\end{figure}
%
\begin{figure}
\caption{Ratio of the calorimeter energy and track momentum
(E/p) for $K_{e3}$ calibration electrons.
The r.m.s. width of the peak is approximately 3\%.
Data (histogram) and Monte Carlo (points) are shown.
The low side tail in the data (which is not well-represented by the Monte
Carlo simulation) is under investigation.}
\label{eoverp}
\end{figure}
%
\begin{figure}
\caption{Mean $E/p$ during the run. Shown in the
plot is the average value of $E/p$ for $K_{e3}$ electrons
as a function of run number.  The period of time spanned by
the run was about nine weeks.}
\label{eoverpvst}
\end{figure}
%
\begin{figure}
\caption{$E/p$ mean and r.m.s. width {\it vs.} electron track momentum.
The upper figure shows the results of fits for the mean $E/p$,
restricted to the region near the
peak,
for $K_{e3}$ electrons.
The apparent shift away from unity is an
artifact of the restriction in the fits to the peak region.
The lower figure shows the
r.m.s. width, again restricted to the region
near the peak.
Data (histogram) and Monte Carlo (points) are shown.}
\label{eoverpvsp}
\end{figure}
%
\begin{figure}
\caption{$\pi^+ \pi^-$ invariant mass distributions for events
satisfying all other analysis cuts.
Events from decays in the upstream and downstream regenerator
beams are plotted separately.  Midway through the run the
T,V hodoscope was removed; data are plotted separately from
the periods before (set 1) and after (set 2) its removal.
Data (histogram) and Monte Carlo (points) are shown.
The data's high-side tails contain
contributions from $\delta$-rays which are not simulated by the Monte
Carlo.}
\label{pipimass}
\end{figure}
%
\begin{figure}
\caption{$p_T^2$ distributions for $K \rightarrow \pi^+ \pi^-$ decays
satisfying all other analysis cuts.
Events from decays in the upstream and downstream regenerator
beams are plotted separately.  Midway through the run the
T,V hodoscope was removed; data are plotted separately from
the periods before (set 1) and after (set 2) its removal.
Data (histogram) and Monte Carlo (points) are shown.
Some smearing of the data's coherent peak is due to the
presence of $\delta$-rays which are not simulated by the Monte
Carlo.}
\label{charpt}
\end{figure}
%
\begin{figure}
\caption{Decay vertex $z$ distributions for $K \rightarrow \pi^+ \pi^-$
decays satisfying all other $Arg(\eta_{+-})$
analysis cuts. To reduce backgrounds associated
with interactions of beam hadrons in material near the downstream
regenerator, events with vertex z between 127~m and 129~m were discarded.
Data (histogram) and Monte Carlo (points) are shown for each beam.
The T,V hodoscope, located near $z=141$~m,
was included in the charged mode trigger
for set 1 data.}
\label{pipiz}
\end{figure}
%
\begin{figure}
\caption{Reconstructed
kaon energy distributions for $K \rightarrow \pi^+ \pi^-$ decays
satisfying all other analysis cuts.
Data (histogram) and Monte Carlo (points) are shown for each beam
in the two data sets.}
\label{pipie}
\end{figure}
%
\begin{figure}
\caption{Photon energy spectrum in the $K$ rest frame for
$\pi^+ \pi^-\gamma$ decays satisfying all other cuts.
Events are required to have $E_\gamma^* > 0.02$ GeV.
Data (histogram) and Monte Carlo (points) are shown for the
entire data sample. The vertical dashed line indicates the location
of the 0.02 GeV cut.}
\label{estargamma}
\end{figure}
%
\begin{figure}
\caption{$\pi^+ \pi^-\gamma$ invariant mass distributions
for events satisfying all other cuts.
Data (shown as histograms) are plotted separately for each beam.
Results of the Monte Carlo
simulation (shown as points)
do not include
background modeling.
}
\label{pipigmass}
\end{figure}
%
\begin{figure}
\caption{$p_T^2$ distributions for $K \rightarrow \pi^+ \pi^-\gamma$ decays
satisfying all other cuts.
Data (histogram) and Monte Carlo (points) are shown,
plotted separately for each beam.
The Monte Carlo did not
simulate
backgrounds.}
\label{pipigptsq}
\end{figure}
%
\begin{figure}
\caption{Decay vertex $z$ distributions for
$K \rightarrow \pi^+ \pi^-\gamma$ decays
satisfying all other cuts.
Data (histogram) and Monte Carlo (points) are shown,
plotted separately for each beam in the two data sets.
}
\label{pipigz}
\end{figure}
%
\begin{figure}
\caption{$K$ energy distributions for
$K \rightarrow \pi^+ \pi^-\gamma$ decays satisfying all other cuts.
Data (histogram) and Monte Carlo (points) are shown.}
\label{pipige}
\end{figure}
%
\begin{figure}
\caption{$\pi^0 \pi^0$ invariant mass distributions
for events satisfying all other cuts.
Data (histogram) and Monte Carlo (points) are shown.  In the plots
on the left side of the figure, the simulation includes both
signal and backgrounds. In the plots on the right, only the
simulated backgrounds are included in the
Monte Carlo sample.}
\label{2pi0mass}
\end{figure}
\begin{figure}
\caption{Illustration of ``ring number" for a
$K \rightarrow \pi^0 \pi^0$ decay. In this example,
an event's center of energy in the calorimeter is at
$(x,y) = (14,4)$~cm, closer to the upper beam which is centered at
$(x,y) = (0,11.6)$~cm.
The square whose perimeter contains the center of energy
has side length 2~$\times$~14~cm and area 784~${\rm cm^2}$.
As a result,
the event's ring number is 784~${\rm cm^2}$.}
\label{ringdiagram}
\end{figure}
%
\begin{figure}
\caption{$\pi^0 \pi^0$ ring distributions
for events satisfying all other cuts.
Data (histogram) and Monte Carlo (points) are shown.  In the plots
on the left side of the figure, the simulation includes both
signal and backgrounds. In the plots on the right, only the
simulated backgrounds are included in the
Monte Carlo sample.}
\label{ringplot}
\end{figure}
%
\begin{figure}
\caption{Decay vertex $z$ distributions for
$K \rightarrow \pi^0 \pi^0$ events
satisfying all other cuts.
Data (histogram) and Monte Carlo (points) are shown.}
\label{p0p0z}
\end{figure}
%
\begin{figure}
\caption{Reconstructed kaon energy distributions for
$K \rightarrow \pi^0 \pi^0$ decays satisfying all other cuts.
Data (histogram) and Monte Carlo (points) are shown.}
\label{p0p0e}
\end{figure}
%
\begin{figure}
\caption{
Angle between the target-to-vertex direction
and the
$z$ axis
in $K_{e3}$ decays,
projected onto the horizontal plane. Wings of the distribution
correspond to $K$ mesons which scattered in beamline elements before
decaying.
Data (histogram) and Monte Carlo (points) are shown.}
\label{elliottbeams}
\end{figure}
%
\begin{figure}
\caption{Comparison of the
data and Monte Carlo $K_{e3}$
electron illuminations at the east edge of
the HDRA aperture in 1~mm bins.
Data (histogram) and Monte Carlo (points) are shown.}
\label{elliotthdra}
\end{figure}
%
\begin{figure}
\caption{$\pi^\pm$
track illuminations at the HDRA.
Data (histogram) and Monte Carlo (points) are shown.}
\label{chgillumhdra}
\end{figure}
%
\begin{figure}
\caption{$\pi^\pm$ track
illuminations at the lead glass array.
Data (histogram) and Monte Carlo (points) are shown.}
\label{chgillumglass}
\end{figure}
%
\begin{figure}
\caption{Decay vertex $z$ distributions for $K \rightarrow 3\pi^0$ decays.
Data (histogram) and Monte Carlo (points) are shown.  The large $3\pi^0$
sample provides a valuable check of our $\pi^0\pi^0$
acceptance calculation.}
\label{pi0pi0pi0z}
\end{figure}
%
\begin{figure}
\caption{Decay vertex $z$ distributions for
$K \rightarrow \pi e \nu$ decays.
Data (histogram) and Monte Carlo (points) are shown.  The large $K_{e3}$
sample allows us to check our $\pi^+\pi^-$ acceptance calculation.}
\label{ke3z}
\end{figure}
%
\begin{figure}
\caption{Magnitude and phase of $\eta_{+-\gamma}$ from this
experiment and from E731. The smaller error bars indicate statistical
uncertainties.  The larger errors correspond to a sum, in quadrature,
of the quoted statistical and systematic errors.
For comparison,
the vertical dashed lines indicate the $\pm 1 \sigma$ bounds
on our $\phi_{+-}$ measurement
while the horizontal dashed lines indicate the
$\pm 1 \sigma$ bounds on the Particle Data Group's world-average
value for $\vert \eta_{+-} \vert$. }
\label{etappgplot}
\end{figure}

\begin{references}
\bibitem{panda}PANDA is an ACP-based data acquisition system
developed by the Online Systems Group at Fermilab.
\end{references}
\end{document}